\documentclass[10pt,a4paper]{article}

\usepackage[T1]{fontenc}
\usepackage[utf8]{inputenc}
\usepackage{lmodern}
\usepackage{microtype}
\usepackage[margin=2.4cm]{geometry}
\usepackage{amsmath,amssymb,amsthm}
\usepackage{graphicx}
\usepackage{booktabs}
\usepackage{setspace}
\usepackage[labelfont=bf,font=footnotesize]{caption}
\usepackage[numbers,sort&compress]{natbib}
\usepackage{xcolor}
\usepackage{xurl}
\usepackage[colorlinks=true,linkcolor=black,citecolor=blue!50!black,
            urlcolor=blue!50!black]{hyperref}

\usepackage{mathtools}
\usepackage{bm}
\usepackage{float}
\usepackage[Symbol]{upgreek}
\usepackage{leftidx}
\usepackage{epstopdf}
\usepackage{units}
\usepackage{multirow}
\usepackage[caption=false]{subfig}
\let\subfigure\subfloat
\graphicspath{{figures/}}

\newcommand{\PaperTitle}{Wave propagation analysis for a second strain gradient rod theory}
\newcommand{\PaperAuthors}{Guang Zhu$^{1}$, Christophe Droz$^{1}$, Abdelmalek Zine$^{2}$ and Mohamed Ichchou$^{1}$}
\newcommand{\PaperAffil}{%
  $^{1}$Vibroacoustics \& Complex Media Research Group, LTDS -- CNRS UMR 5513, Centre Lyonnais d'Acoustique CeLyA, \'Ecole Centrale de Lyon, Université de Lyon, France\\
$^{2}$Institute Camille Jordan -- CNRS UMR 5208, École Centrale de Lyon, France}
\newcommand{\PaperDate}{Accepted manuscript, originally 2019}
\newcommand{\PaperKeywords}{Dynamic behavior, Energy flow, Enriched rod model, Reflection and transmission, Second strain gradient theory, Wave propagation features}

\newcommand{\PaperDOI}{10.1016/j.cja.2019.10.006}
\newcommand{\HalID}{hal-02712551}
\newcommand{\ArxivID}{}          % ex. 2609.01234 --- laisser vide si inconnu

\newcommand{\CiteAs}{%
  G. Zhu, C. Droz, A. Zine, M. Ichchou, \emph{\PaperTitle}, Chinese Journal of Aeronautics 33 (2020) 2563-2574, doi:10.1016/j.cja.2019.10.006}
\makeatletter
\newcommand{\HasDOI}{%
  \ifx\PaperDOI\@empty\expandafter\@secondoftwo\else\expandafter\@firstoftwo\fi}
\newcommand{\HasArxiv}{%
  \ifx\ArxivID\@empty\expandafter\@secondoftwo\else\expandafter\@firstoftwo\fi}
\makeatother

\providecommand{\doi}[1]{\href{https://doi.org/#1}{doi:#1}} 

\begin{document}

% Title page: bold title; authors; copyable citation.
% Affiliations, date, HAL/DOI/AAM live in footnotes (fnsymbol, then arabic resumes).
\begingroup
\renewcommand{\thefootnote}{\fnsymbol{footnote}}
\begin{center}
{\LARGE\bfseries \PaperTitle\par}
\vspace{0.7em}
{\normalsize \PaperAuthors\footnotemark[1]\footnotemark[2]\par}
\end{center}
\footnotetext[1]{\mbox{}\\[-0.7em]\raggedright\PaperAffil}
\footnotetext[2]{\PaperDate.
  Author's version with AI-assisted typesetting and layout. Not the publisher version of record.
  \HasDOI{Definitive publication: \doi{\PaperDOI}.}{No journal DOI yet.}
  HAL: \href{https://hal.science/\HalID}{\HalID}.
  \HasArxiv{arXiv: \href{https://arxiv.org/abs/\ArxivID}{\ArxivID}.}{}}
\endgroup
\setcounter{footnote}{0}

\vspace{0.85em}
{\small
\noindent\textit{How to cite this version.}\\[0.35em]
\raggedright \CiteAs
\par}
\vspace{1.05em}

\begin{abstract}
\noindent
In this work, an enriched model describing the longitudinal wave propagation is established based on Mindlin's Second Strain Gradient (SSG) theory, which can describe the heterogeneity caused by the micro-structure interactions in the frame of continuum mechanics. The governing equation and associated boundary conditions are derived based on Hamilton's principle, then the dispersion relation of non-classical longitudinal wave together with the extra-waves appearing exclusively in SSG theory model are investigated. The investigations are based on the modal density, energy flow, and forced response of the rod. Wave transmission and reflection through planar interfaces based on the proposed model have been calculated. Finally, the results of the enriched model are well interpreted by comparing with the classical theory results, and some useful conclusions are derived on the SSG theory based model in the wave propagation characterization.
\end{abstract}


% ============================================================
% FILL — corps : corps du papier
% ============================================================

\section{Introduction}
\label{Introduction}
In conventional continuum mechanics theory, the strain energy density for the material is assumed to depend only on the classical strain, and the material is modeled to be a continuous mass rather than as discrete particles. In reality, no material is an ideal continuum. Both natural and man-made materials have a complicated internal structure characterized by micro-structural details. Hence, it is not possible to describe the more complex interactions occurring in generalized continua by means of the sole Cauchy stress tensor.\cite{r1} This is particularly true for the long-range interactions and micro-structural deformations, which leads to the failure of conventional theory in many situations.\cite{r2} One solution is to describe each micro-particle separately with classical continuum theory as in atomistic models, but it requires enormous time and computational resource, moreover, the numerical simulation of large structures with high mechanical and geometrical contrasts is also facing to ill-conditioned problems.

To bridge the gap between atomistic models and classical continuum mechanics, a number of theories and approaches have been developed as extensions of the conventional continuum mechanics. Among them, `Couple Stress theory' (CS theory) was firstly suggested in 1962 by Toupin et al.\cite{r3,r4} In CS theory, the gradient of the rotation vector is introduced into the strain energy density function. Subsequently, the `Strain Gradient theory' (SG theory) was developed by introducing all the components of strain gradient into the strain energy density function.\cite{r5} In 1965, Mindlin\cite{r6} established the `Second Strain Gradient theory' (SSG theory), in which the strain energy density is considered to depend on the second and third derivatives of the displacement, along with the classical strain tensor. In 2002, Modified Couple Stress (MCS) theory is developed with an additional equilibrium relation to govern the behavior of the `couple'.\cite{r7,r8} Another approach to describe the micro-structural deformation effect is by increasing the degrees of freedom of the particles.\cite{r9,r10,r11} In this category, the micropolar theory and the micro-stretch theory, investigated by Eringen \cite{r12,r13}, a director vector and a rigid vector representing the inner rotation are introduced. Generalized elasticity theory formulations enrich the classical elasticity theory by means of additional higher order spatial derivatives of relevant state variables (e.g. strains, stresses). These higher order terms are usually accompanied with additional length scale parameters (or higher order constants) in order to measure the scale effects which results from the underlying micro-structure interactions.\cite{r14}

The generalized continuum theories already have some successful applications, namely, being capable of capturing the scale effects of experimentally observed mechanics behaviors where the specimen characteristic length or the wavelength of the disturbance are comparable to the lengths of the micro-structure in the media.\cite{r15} To be specific, the enriched model can be developed with the utilization of Hamilton principle, and the derived models are employed to investigate a broad range of problems related to various beam and plate structures.\cite{r16,r17,r18} One can cite the static deflection and bending stiffness analyses,\cite{r19,r20} fracture behaviors,\cite{r21,r22} structural free vibration and forced vibration,\cite{r23,r24,r25} nonlinear deformation and nonlinear dynamic performances,\cite{r26,r27,r28} flexoelectricity property\cite{r29} and functionally graded structure property analysis\cite{r30,r31}, etc.

Furthermore, these generalized continuum theories introduce different deformation behaviors for the particles and also more complex modal solutions. A known consequence is the existence of new wave modes which could not be observed in the classical elastic solids. Recent work by Suiker et al.\cite{r31,r32} predicted `dispersive' body waves based on the proposed second gradient micro-polar formulation. When the wavelength reaches the order of the particle size, the dispersion becomes more prominent. Wave propagation characteristics are also investigated by Gopalakrishnan\cite{r33} using Eringen's Stress Gradient model and Mindlin's Strain Gradient model, as well as by Li et al.\cite{r34} with nonlocal strain gradient theory. Atomistic length scale parameters are brought into the continuum governing equations. These scale parameters are proven to significantly affect the wave propagation referring to not only the dispersion relation, but also the escape frequency, phase speeds and group speeds in the structures, and the prediction results\cite{r34} match very well with those predicted using atomistic simulation.

The micro-structure effect not only lead to different body wave modes in the solid but also cause different waves reflection and transmission on a discontinuous surface of generalized continua. In the context of generalized continua, Placidi et al.\cite{r35} studied compression and shear wave propagation in second gradient continua. They noticed that the effect of the second gradient parameter are important for the reflection and transmission coefficients at surfaces of discontinuity. Dell'Isola et al.\cite{r36} deduced the mechanical energy equilibrium for second gradient material, based on which they estimated that the reflection and transmission coefficients at plane displacement discontinuity surfaces are significantly frequency-dependent. According to the research of Li et al.,\cite{r37,r38} the reflection and transmission coefficients at the interface of two generalized medias are not only dependent of the microstructure's parameters, but also of the incident angular frequency. This phenomenon is owing to the dispersive nature of the reflected and transmitted waves, and also to the additional surface waves resulting from the micro-structure effect, and this phenomenon only becomes pronounced when the incident wavelength is close to the characteristic length of the micro-structure.

Mindlin's SSG theory is considered as one of the most effective gradient elasticity theories. Even though the lack of available experimental measurement techniques for second gradient elastic moduli and being short of the physical interpretation of the higher order constants yield some criticisms, one can note that the CS theory and SG theory are particular cases of the SSG theory with less high order constants. Therefore, this theory is more general than the CS and SG theories. Furthermore, the research works involving both the SSG theory and wave propagation are seldom in the published literature. The most detailed wave propagation features such as the modal density, the energy flux, have not been discussed, and the reflection and transmission waves based on SSG theory at an interface have not reported so far.

To study the wave propagation features of rod structure having heterogeneity caused by inner micro-structure interactions, in the meantime avoid the enormous time and computational resource that atomistic simulation method requires, in the present work, we resort to SSG theory to describe the heterogeneity caused by the micro-structure interactions in the frame of continuum mechanics. The governing equation and associated boundary conditions for the enriched model are derived based on Hamilton's principle in Section~\ref{sec:3.1}. Then the enriched wave propagation features of the structure are studied in Section~\ref{sec:3.2}. Investigations are then conducted on the modal density (Section~\ref{sec:3.3}), the energy flow (Section~\ref{sec:3.4}), and the forced response of the rod (Section~\ref{sec:3.5}). In Section~\ref{sec:4}, the wave transmission and reflection at a planar surface between two waveguides of second gradient material are analyzed for the considered model. Numerical applications of the above are discussed in Section~\ref{sec:5} and conclusions are drawn in Section~\ref{sec:6}. This work not only aims to contribute to the analysis of wave propagation in SSG theory based media, but also aims at providing an original solution to analyze the wave dispersion characteristics of complex media with micro-scale periodicity.

\section{Overview of the SSG theory}
\label{sec:2}
According to Mindlin's research,\cite{r6} in SSG theory, the potential energy density $\bar{u}$ is assumed to be a function of three tensors $\boldsymbol{\varepsilon}$, $\boldsymbol{\eta}$ and $\boldsymbol{\xi}$, in which $\boldsymbol{\varepsilon}$ is the classical symmetric strain tensor, $\boldsymbol{\eta}$ and $\boldsymbol{\xi}$ are respectively the second gradient of the displacement vector and the third gradient of the displacement vector,
\[
\begin{cases}
\varepsilon = \frac{1}{2}\left(\nabla\boldsymbol{u} + \boldsymbol{u}\nabla\right)\\
\eta = \nabla\nabla u\\
\xi = \nabla\nabla\nabla u
\end{cases}
\]
Since the media is composed of numerous crystals with random orientation, the material can be considered homogeneous and isotropic, then strain energy density of the material is expressed as
\begin{equation}
\begin{split}
\bar{u} ={}& \frac{1}{2}\lambda\varepsilon_{ii}\varepsilon_{jj} + \mu\varepsilon_{ij}\varepsilon_{ij} + a_1\eta_{ijj}\eta_{ikk} + a_2\eta_{iik}\eta_{kjj} + a_3\eta_{iik}\eta_{jjk} + a_4\eta_{ijk}\eta_{ijk}\\
&+ a_5\eta_{ijk}\eta_{kji} + b_1\xi_{iijj}\xi_{kkll} + b_2\xi_{ijkk}\xi_{ijll} + b_3\xi_{iijk}\xi_{jkll} + b_4\xi_{iijk}\xi_{llkj} + b_5\xi_{iijk}\xi_{lljk}\\
&+ b_6\xi_{ijkl}\xi_{ijkl} + b_7\xi_{ijkl}\xi_{jkli} + c_1\varepsilon_{ii}\xi_{jjkk} + c_2\varepsilon_{ij}\xi_{ijkk} + c_3\varepsilon_{ij}\xi_{kkij}
\end{split}
\label{eq:1}
\end{equation}
where $\varepsilon_{ij}$, $\eta_{ijk}$ and $\xi_{ijkl}(i,j,k,l=1,2,3)$ are the components of tensors $\boldsymbol{\varepsilon}$, $\boldsymbol{\eta}$ and $\boldsymbol{\xi}$, their expressions can be written in terms of the components of displacement vector as
\begin{equation}
\begin{cases}
\varepsilon_{ij} = \frac{1}{2}\left(\frac{\partial u_j}{\partial x_i} + \frac{\partial u_i}{\partial x_j}\right) = \frac{1}{2}\left(\partial_j u_i + \partial_i u_j\right)\\
\eta_{ijk} = \frac{\partial^2 u_k}{\partial x_i\partial x_j} = \partial_i\partial_j u_k\\
\xi_{ijkl} = \frac{\partial^3 u_l}{\partial x_i\partial x_j\partial x_k} = \partial_i\partial_j\partial_k u_l
\end{cases}
\label{eq:2}
\end{equation}
$\lambda$ and $\mu$ are the usual Lame's constants. Parameters $a_i$, $b_i$ and $c_i$ are higher order material constants which particularly appear in SSG theory. If the potential energy density is assumed to depend on the strain and the gradient of the strain, setting $b_i=0$ and $c_i=0$, then it results the strain gradient theory. The conjugate stresses are defined by differentiating the strain energy density $\bar{u}$ to the corresponding strain components,
\begin{equation}
\begin{cases}
\boldsymbol{\sigma} = \frac{\partial\bar{u}}{\partial\boldsymbol{\varepsilon}}\\
\boldsymbol{\tau} = \frac{\partial\bar{u}}{\partial\boldsymbol{\eta}}\\
\boldsymbol{\pi} = \frac{\partial\bar{u}}{\partial\boldsymbol{\xi}}
\end{cases}
\label{eq:3}
\end{equation}
It should be noted here, even though the stress tensors $\boldsymbol{\sigma}$, $\boldsymbol{\tau}$ and $\boldsymbol{\pi}$ are respectively named with the classical stress, the first higher order stress and the second higher order stress, their values could be related with all three order strains as,\cite{r6}
\[
\begin{cases}
\sigma_{pq} = \lambda\varepsilon_{ii}\delta_{pq} + 2\mu\varepsilon_{pq} + c_1\xi_{iijj}\delta_{pq} + c_2\xi_{pqii} + \frac{1}{2}c_3\left(\xi_{iipq} + \xi_{iiqp}\right)\\
\tau_{pqr} = a_1\left(\eta_{pii}\delta_{qr} + \eta_{qii}\delta_{pr}\right) + \frac{1}{2}a_2\left(\eta_{iip}\delta_{qr} + 2\eta_{rii}\delta_{qp} + \eta_{iiq}\delta_{pr}\right)\\
+2a_3\eta_{iir}\delta_{pq} + 2a_4\eta_{pqr} + a_5\left(\eta_{rqp} + \eta_{rpq}\right)\\
\pi_{pqrs} = \frac{2}{3}b_1\xi_{iijj}\delta_{pqrs} + \frac{2}{3}b_2\xi_{jkii}\delta_{jkpqrs} + \frac{1}{6}b_3\left[\left(\xi_{iijk} + \xi_{iikj}\right)\delta_{jkpqrs} + 2\xi_{jsii}\delta_{jpqr}\right]\\
+\frac{2}{3}b_4\xi_{iisj}\delta_{jpqr} + \frac{2}{3}b_5\xi_{iijs}\delta_{ipqr} + 2b_6\xi_{pqrs} + \frac{2}{3}b_7\left(\xi_{qrsp} + \xi_{rspq} + \xi_{spqr}\right)\\
+\frac{1}{3}c_1\varepsilon_{ii}\delta_{pqrs} + \frac{1}{3}c_2\varepsilon_{ij}\delta_{ijpqrs} + \frac{1}{3}c_3\varepsilon_{is}\delta_{ipqr}
\end{cases}
\]
in which $\delta_{ij}$ is the Kronecker delta function and
\[
\delta_{pqrs} = \delta_{ij}\delta_{kl} + \delta_{ik}\delta_{jl} + \delta_{jk}\delta_{il},\delta_{ijklmn} = \delta_{ik}\delta_{jl}\delta_{mn} + \delta_{ik}\delta_{jm}\delta_{ln} + \delta_{il}\delta_{jm}\delta_{kn}
\]

\section{Analysis of wave dispersion characteristics, energy flow and modal density}
\label{sec:3}

\subsection{The derivation of governing equation and boundary conditions}
\label{sec:3.1}
\begin{figure}[H]
\centering
\includegraphics[width=0.75\linewidth]{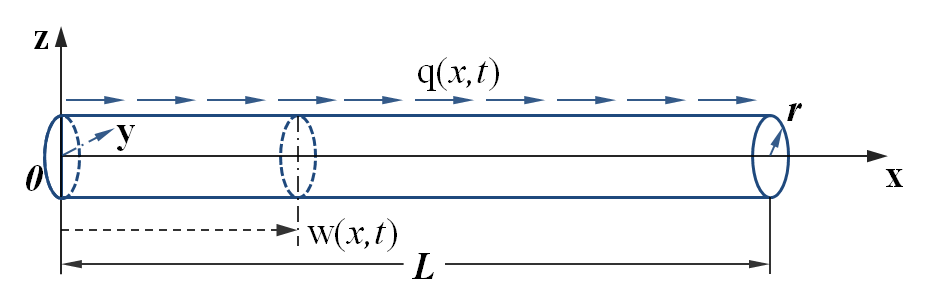}
\caption{Coordinate system and kinematic parameters of rod.}
\label{fig:1}
\end{figure}

The multi-scale modeling begins with establishing the motion equation of the rod for free vibration. The coordinate system and kinematic parameters of the model are illustrated in Fig.~\ref{fig:1}. The rod is assumed to be uniform, homogeneous and initially straight along the $x$-direction with length $L$. $r$ denotes the radius of the circular cross section, and $q(x,t)$ denotes the resultant of external forces acting on the rod in $x$ direction as force per unit axial length.

The displacement field in a rod model is written as,
\begin{equation}
\begin{cases}
u_x(x,y,z,t) = w(x,t)\\
u_y(x,y,z,t) = 0\\
u_z(x,y,z,t) = 0
\end{cases}
\label{eq:4}
\end{equation}
where $u_x$, $u_y$, and $u_z$ denote the displacement components of the rod particles along $x$, $y$, and $z$ directions respectively. According to Eq.~(\ref{eq:2}), the nonzero components of the strain tensor $\boldsymbol{\varepsilon},\boldsymbol{\eta},\text{and},\boldsymbol{\xi}$ can be depicted as
\begin{equation}
\begin{cases}
\varepsilon_{11} = \frac{\partial w}{\partial x}\\
\eta_{111} = \frac{\partial^2 w}{\partial x^2}\\
\xi_{1111} = \frac{\partial^3 w}{\partial x^3}
\end{cases}
\label{eq:5}
\end{equation}
Substituting the nonzero components of $\boldsymbol{\varepsilon},\boldsymbol{\eta},\text{and},\boldsymbol{\xi}$, as mentioned in Eq.~(\ref{eq:5}) into Eq.~(1), the strain energy density $\bar{u}$ for the SSG theory rod can be simplified as
\begin{equation}
\begin{split}
\bar{u} ={}& \frac{1}{2}\lambda{\varepsilon_{11}}^2 + \mu{\varepsilon_{11}}^2 + (a_1 + a_2 + a_3 + a_4 + a_5){\eta_{111}}^2\\
&+(b_1 + b_2 + b_3 + b_4 + b_5 + b_6 + b_7){\xi_{1111}}^2\\
&+(c_1 + c_2 + c_3)\varepsilon_{11}\xi_{1111}
\end{split}
\label{eq:6}
\end{equation}
Lame's constants $\lambda$ and $\mu$ are related to the Young's modulus $E$ and the Poisson's ratio $\nu$, as $\lambda = \frac{\nu E}{(1+\nu)(1-2\nu)}$, so the Lame's constants here can be determined with experimentally tested $E$ and $\nu$. However, there are still not yet standard experiments to determine the higher order material constants $a_i$, $b_i$, and $c_i$, so many scientists and engineers resorted to atomistic simulation method. In present model, the higher order material constants refer to the article by Shodja et al. (2012)\cite{r39} with atomistic approach. The strain energy of the rod is calculated by integrating its density over its volume. Combination of Eq.~(\ref{eq:5}) and Eq.~(\ref{eq:6}), the strain energy of the SSG theory based rod model can be simplified as
\begin{equation}
\begin{split}
U ={}& \int_0^L\int_A \bar{u}\,\mathrm{d}A\mathrm{d}x = \frac{1}{2}\int_0^L\left[\chi EA\left(\frac{\partial w}{\partial x}\right)^2 + B_1\left(\frac{\partial^2 w}{\partial x^2}\right)^2 + B_2\left(\frac{\partial^3 w}{\partial x^3}\right)^2\right.\\
&\left.+B_3\frac{\partial w}{\partial x}\cdot\frac{\partial^3 w}{\partial x^3}\right]\mathrm{d}x
\end{split}
\label{eq:7}
\end{equation}
in which,
\[
\begin{cases}
B_1 = 2A(a_1 + a_2 + a_3 + a_4 + a_5)\\
B_2 = 2A(b_1 + b_2 + b_3 + b_4 + b_5 + b_6 + b_7)\\
B_3 = 2A(c_1 + c_2 + c_3)
\end{cases}
\]
with parameter $A$ as the area of the rod cross section, and $\chi = \frac{1-\nu}{(1+\nu)(1-2\nu)}$. To avoid the Poisson effect, the results in the present work will be calculated with setting $\chi = 1$.
From the displacement field represented in Eq.~(\ref{eq:4}), the kinetic energy of the rod can be written as,
\begin{equation}
\textstyle T = \frac{1}{2}\int_0^L\int_A\rho\left(\frac{\partial w}{\partial t}\right)^2\mathrm{d}A\mathrm{d}x = \frac{1}{2}\int_0^L m_0\left(\frac{\partial w}{\partial t}\right)^2\mathrm{d}x
\label{eq:8}
\end{equation}
where $m_0 = \rho A$,$\rho$ denotes density.

The virtual work done by available loads on the rod, including body loads and boundary surface loads, during occurrence a variation in the geometrical state of the beam at given time $t>0$ is expressed as
\begin{equation}
\textstyle \delta W = \int_0^L(q\delta w)\mathrm{d}x + \left[P_0\delta w + P_1\delta\left(\frac{\partial w}{\partial x}\right) + P_2\delta\left(\frac{\partial^2 w}{\partial x^2}\right)\right]\Bigg|\begin{matrix} x=L\\ x=0\end{matrix}
\label{eq:9}
\end{equation}
in which $P_i$ are the end-sectional loads dual to corresponding kinematic parameters in performing work, and these loads are resultants of not only classical stresses but also higher order stresses on the surface of an end section.

Until here, the strain energy, the kinetic energy and the virtual work done by available loads for the SSG theory rod have been expressed out. To derive the motion equation and the associated boundary conditions, the Hamilton principle is employed as
\begin{equation}
\textstyle \int_{t_1}^{t_2}(\delta T - \delta U + \delta W)\quad\mathrm{d}t = 0
\label{eq:10}
\end{equation}
where $\delta T$ and $\delta U$ are the variation of kinetic energy and the variation of strain energy which both have been described in previous formulas. Substituting Eq.~(\ref{eq:7}), Eq.~(\ref{eq:8}) and Eq.~(\ref{eq:9}) into Eq.~(\ref{eq:10}), and performing the indicated variations, the governing equation of the enriched model is obtained as
\begin{equation}
m_0\frac{\partial^2 w}{\partial t^2}
= EA\frac{\partial^2 w}{\partial x^2}
+(B_3-B_1)\frac{\partial^4 w}{\partial x^4}
+B_2\frac{\partial^6 w}{\partial x^6}
+q
\label{eq:11}
\end{equation}
(with \(\chi=1\)) and the associated boundary conditions at the rod ends (\(x=0,L\)) as
\begin{equation}
\begin{cases}
P_0 = EA\frac{\partial w}{\partial x} + (B_3 - B_1)\frac{\partial^3 w}{\partial x^3} + B_2\frac{\partial^5 w}{\partial x^5}\ \text{or}\ \delta w = 0\\
P_1 = \left(B_1 - \frac{B_3}{2}\right)\frac{\partial^2 w}{\partial x^2} - B_2\frac{\partial^4 w}{\partial x^4}\ \text{or}\ \delta\left(\frac{\partial w}{\partial x}\right) = 0\\
P_2 = \frac{B_3}{2}\cdot\frac{\partial w}{\partial x} + B_2\frac{\partial^3 w}{\partial x^3}\ \text{or}\ \delta\left(\frac{\partial^2 w}{\partial x^2}\right) = 0
\end{cases}
\label{eq:12}
\end{equation}
In each boundary condition, there are two kinds: a loading type or a geometric type, only one type is required to be considered at an end-section of the rod in each case. From the expressions of the governing equation and boundary conditions, we can also see if we set the higher order material constants $a_i=0$, $b_i=0$ and $c_i=0$, the classical theory results can be achieved.\cite{r41}

For further exploration and parametric study of the enriched mechanical properties of SSG theory based rod, normalization of the governing equations and the associated boundary conditions is necessary, so the following normalization parameters are defined,
\begin{equation}
\begin{cases}
\bar{x} = \frac{x}{L}\\
\bar{w} = \frac{w}{L}\\
\bar{\omega} = \frac{\omega}{\omega_k}\\
\tau = \frac{t}{t_k}
\end{cases}
\label{eq:13}
\end{equation}
the term $\omega_k$ is defined as
\[
\omega_k = \sqrt{\frac{E}{\rho}}\cdot\frac{\kappa}{L}
\]
which is the normalization parameter for angular frequency $\omega$. In this case, it is set to be one natural frequency of the structure. $\kappa$ is a constant determined by the boundary conditions of the model. In the presented model, $\kappa$ is set to be $\pi$. The normalization parameter for time $t$ is expressed as
\[
t_k = \frac{1}{\omega_k} = \sqrt{\frac{\rho}{E}}\cdot\frac{L}{\kappa}
\]
Substituting the newly defined variables into the governing equation and boundary conditions yields the normalized model
\begin{equation}
\frac{E}{\mu}\frac{\partial^2\bar{w}}{\partial\bar{x}^2}
+\frac{(B_3-B_1)}{\mu AL^2}\frac{\partial^4\bar{w}}{\partial\bar{x}^4}
+\frac{B_2}{\mu AL^4}\frac{\partial^6\bar{w}}{\partial\bar{x}^6}
+\bar{q}
-\frac{m_0}{\mu A}\frac{L^2}{t_k^2}\frac{\partial^2\bar{w}}{\partial\tau^2}
=0
\label{eq:14}
\end{equation}
in which \(\bar{q}=qL/(\mu A)\) is the dimensionless body-load intensity. (Equivalently, because \(t_k^{-2}=(E/\rho)(\kappa/L)^2\) and \(m_0=\rho A\), the coefficient of the inertia term may be written \(\frac{E}{\mu}\kappa^2\).)
\begin{equation}
\begin{cases}
\bar{P}_0 = \frac{E}{\mu}\cdot\frac{\partial\bar{w}}{\partial\bar{x}} + \frac{B_3 - B_1}{\mu AL^2}\cdot\frac{\partial^3\bar{w}}{\partial\bar{x}^3} + \frac{B_2}{\mu AL^4}\cdot\frac{\partial^5\bar{w}}{\partial\bar{x}^5}\ \text{or}\ \delta\bar{w} = 0\\
\bar{P}_1 = \frac{2B_1 - B_3}{2\mu AL^2}\cdot\frac{\partial^2\bar{w}}{\partial\bar{x}^2} - \frac{B_2}{\mu AL^4}\cdot\frac{\partial^4\bar{w}}{\partial\bar{x}^4}\ \text{or}\ \delta\left(\frac{\partial\bar{w}}{\partial\bar{x}}\right) = 0\\
\bar{P}_2 = \frac{B_3}{2\mu AL^2}\cdot\frac{\partial\bar{w}}{\partial\bar{x}} + \frac{B_2}{\mu AL^4}\cdot\frac{\partial^3\bar{w}}{\partial\bar{x}^3}\ \text{or}\ \delta\left(\frac{\partial^2\bar{w}}{\partial\bar{x}^2}\right) = 0
\end{cases}
\label{eq:15}
\end{equation}

\subsection{Dispersion relation}
\label{sec:3.2}
For the material with heterogeneity due to micro-structures effects, long range interaction has great influences on the deformation process, especially when the wavelength is of the same order with the heterogeneity, and the classical elasticity theory is clearly not capable of describing these physical properties. In the following section, the dispersion relation of the waves propagating in the rod structure will be investigated based on the proposed model.

Assuming the external loading $\bar{q}=0$, the free wave propagation modes in the proposed enriched model can be achieved by injecting the general exponential form of wave propagation solution:
\begin{equation}
w = w_0\mathrm{e}^{\mathrm{i}(\omega t - kx)}
\label{eq:16}
\end{equation}
into the governing equation, one 6th-order dispersion relation can be obtained as
\begin{equation}
EAk^2 - (B_3 - B_1)k^4 + B_2k^6 - m_0\omega^2 = 0
\label{eq:17}
\end{equation}
For parametric study, wavenumber $k$ should also be normalized with $\bar{k} = kL$, then the normalized dispersion relation is shown
\begin{equation}
\textstyle \frac{B_2}{L^4}\bar{k}^6 - \frac{B_3 - B_1}{L^2}\bar{k}^4 + EA\bar{k}^2 - \frac{m_0\bar{\omega}^2L^2}{t_k^2} = 0
\label{eq:18}
\end{equation}
According to the form of the dispersion relation, we can predict that three different modes can be generated in second strain gradient elastic solid.

\subsection{Modal density}
\label{sec:3.3}
When a structure or a system is subject to excitation of high frequencies, its response may involve a large number of high-order modes. Meanwhile in high frequency range, wavelength tends to be more comparable to the length of inner micro-structure, and its influence on wave propagation becomes more pronounced. The modeling of such systems leads the problem of `high frequency' vibration analysis, and Statistical Energy Analysis (SEA) is one of the most common methods for high frequency vibration analysis. In this section, one important parameter in SEA method, namely `modal density', is studied based on the proposed model. Modal density is one statistics based measurement representing the distribution of modal natural frequencies in the frequency domain. Modal density of a structure at frequency $\omega$ is denoted as $n(\omega)$, which means the number of resonant frequencies in a neighborhood $\Delta\omega$ local to that frequency. It may also be interpreted as the expected number of natural frequencies per radian per second as
\begin{equation}
\textstyle n(\omega) = \frac{dN}{d\omega}
\label{eq:19}
\end{equation}
in which the mode count $N$ represents the number of resonant frequencies below that given frequency $\omega$. The propagating wave at any frequency is characterized by the wavenumber $k$ ($=2\pi/\text{wavelength}$), which can be interpreted as the phase difference per unit distance in wave propagation direction. As the wave propagates from the left-end of the rod to the right-hand end, the total phase change is $kL$. Then wave is reflected by the right-hand end, and back towards the left-hand end. Phase difference will be introduced by the Wave reflection at each end, and that difference varies with different boundary conditions. Assuming $\gamma$ represent the total phase change introduced by reflecting boundaries, the total phase changes as the wave travels one complete circuit around the structure can be expressed as $2kL\pm\gamma$. The ``phase-closure principle'' states that if the total phase change is an integral number of $2\pi's$, the condition is satisfied for a natural mode, and the frequency at which it occurs is a natural frequency of the system. This principle can be written as
\[
2kL\pm\gamma = N\cdot 2\pi
\]
Then we have
\[
N(k) = \frac{kL\pm\gamma/2}{\pi}
\]
As the mode order increases, the number of resonances $N(k)$ in frequency domain become increasingly less sensitive to the boundary conditions. Hence we may consider the mode count as,
\begin{equation}
\textstyle N(k) = \frac{kL}{\pi}
\label{eq:20}
\end{equation}
so the Eq.~(\ref{eq:19}) can be simplified as
\begin{equation}
\textstyle n(\omega) = \frac{\mathrm{d}N}{\mathrm{d}\omega} = \frac{\mathrm{d}N}{\mathrm{d}k}\cdot\frac{\mathrm{dk}}{\mathrm{d}\omega} = \frac{\mathrm{L}}{\pi}\cdot\frac{1}{\mathrm{c_g}}
\label{eq:21}
\end{equation}
in which, the term $c_{\mathrm{g}}$ denotes the group velocity of the studied wave mode
\[
c_{\mathrm{g}} = \frac{\mathrm{d}\omega}{\mathrm{d}k}
\]

\subsection{Energy flux analysis}
\label{sec:3.4}
In this section, the approach to predict the energy velocity at one point in the SSG theory based rod structure is presented. At one observation point $M$, the instantaneous kinetic energy density $E_k(M,t)$, potential energy density $E_s(M,t)$ and the instantaneous active energy flow $\boldsymbol{E_p}(M,t)$ based on the classical elasticity theory are defined as\cite{r42,r43}
\begin{equation}
\begin{cases}
E_k = \frac{\rho}{2}Re(\boldsymbol{V})\cdot Re(\boldsymbol{V})\\
E_s = \frac{1}{2}Re(\boldsymbol{\sigma})\cdot Re(\boldsymbol{\varepsilon})\\
\boldsymbol{E_p} = -Re(\boldsymbol{\sigma})\cdot Re(\boldsymbol{V})
\end{cases}
\label{eq:22}
\end{equation}
where $\rho$ is the mass density, $\boldsymbol{V}$ is the velocity vector, and $\boldsymbol{\sigma}$ and $\boldsymbol{\varepsilon}$ are the classical stress and strain tensors respectively. As the strain and stress tensor are defined differently in the SSG theory, potential energy density is enriched with the higher order components as\cite{r6}
\begin{equation}
E_s = \frac{1}{2}[Re(\boldsymbol{\sigma})\cdot Re(\boldsymbol{\varepsilon}) + Re(\boldsymbol{\tau})\cdot Re(\boldsymbol{\eta}) + Re(\boldsymbol{\pi})\cdot Re(\boldsymbol{\xi})]
\label{eq:23}
\end{equation}
where $\boldsymbol{\tau}$ and $\boldsymbol{\eta}$ are the first higher order strain and stress tensor, $\boldsymbol{\pi}$ and $\boldsymbol{\xi}$ are second higher order strain and stress tensor. The sum of the kinetic and potential energy density yields the instantaneous total energy density:
\begin{equation}
E_{\mathrm{total}} = E_k + E_s
\label{eq:24}
\end{equation}
In the following investigations, the time will be removed by time-averaging. A physical quantity $H$, which represents here an energy or a power density, can generally be expressed as
\[
H = f\cdot g
\]
where $f$ and $g$ are complex harmonic physical variables which denote stress, strain, or displacement in the formulation. Thus the time averaged of $H$ is given by
\[
\langle H\rangle = \frac{\omega}{2\pi}\int\limits_0^{2\pi}\mathrm{Re}(f)\cdot\mathrm{Re}(g)dt = \frac{1}{2}\mathrm{Re}(f\cdot g^*)
\]
where `*' denotes conjugation. For the classical theory, the time averaged kinetic energy, potential energy and active energy flux through the observed cross section $A$ are given as,
\begin{equation}
\begin{cases}
\langle T\rangle = \frac{\rho}{4}\int_A Re(\boldsymbol{V}\cdot\boldsymbol{V}^*)\mathrm{d}A\\
\langle U\rangle = \frac{1}{4}\int_A Re(\boldsymbol{\sigma}\cdot\boldsymbol{\varepsilon}^*)dA\\
\langle\boldsymbol{P}\rangle = -\frac{1}{2}\int_A Re(\boldsymbol{\sigma}\cdot\boldsymbol{V}^*)dA
\end{cases}
\label{eq:25}
\end{equation}
For the SSG theory, the time averaged potential energy of the observed cross section $A$ is given as,
\begin{equation}
\langle U\rangle = \frac{1}{4}\int\limits_A Re(\boldsymbol{\sigma}\cdot\boldsymbol{\varepsilon}^* + \boldsymbol{\tau}\cdot\boldsymbol{\eta}^* + \boldsymbol{\pi}\cdot\boldsymbol{\xi}^*)\mathrm{d}A
\label{eq:26}
\end{equation}
Combined with Eq.~(\ref{eq:8}) and Eq.~(\ref{eq:6}), the time-averaged kinetic energy and the enriched time-averaged potential energy of the studied rod, evaluated on an arbitrary cross-section, are
\begin{equation}
\begin{cases}
\langle T\rangle
=\dfrac{\rho}{4}\displaystyle\int_A\operatorname{Re}(\dot{w}\dot{w}^*)\,\mathrm{d}A
=\dfrac{m_0}{4}\operatorname{Re}(\dot{w}\dot{w}^*),\\[1em]
\langle U\rangle
=\dfrac{1}{4}\operatorname{Re}\Biggl[
EA\Bigl(\dfrac{\partial w}{\partial x}\Bigr)\Bigl(\dfrac{\partial w}{\partial x}\Bigr)^*
+B_1\Bigl(\dfrac{\partial^2 w}{\partial x^2}\Bigr)\Bigl(\dfrac{\partial^2 w}{\partial x^2}\Bigr)^*\\
\qquad
+B_2\Bigl(\dfrac{\partial^3 w}{\partial x^3}\Bigr)\Bigl(\dfrac{\partial^3 w}{\partial x^3}\Bigr)^*
+B_3\Bigl(\dfrac{\partial w}{\partial x}\Bigr)\Bigl(\dfrac{\partial^3 w}{\partial x^3}\Bigr)^*
\Biggr]
\end{cases}
\label{eq:27}
\end{equation}
When it comes to energy flow, the higher order stress should be taken into account. Based on the definition of virtual work, the expression of active energy flux in $x$ direction through the observation section is assumed to be as
\begin{equation}
\textstyle \langle P\rangle = -\frac{1}{2}Re\left[P_0\cdot\dot{w}^* + P_1\cdot\left(\frac{\partial\dot{w}}{\partial x}\right)^* + P_2\cdot\left(\frac{\partial^2\dot{w}}{\partial x^2}\right)^*\right]
\label{eq:28}
\end{equation}
where $P_0$, $P_1$, and $P_2$ are the loads dual to the corresponding kinematic parameters, and `$\dot{w}$' denotes the time derivative of the displacement $w$.
\[
\begin{cases}
P_0 = EA\frac{\partial w}{\partial x} + (B_3 - B_1)\frac{\partial^3 w}{\partial x^3} + B_2\frac{\partial^5 w}{\partial x^5}\\
P_1 = \left(B_1 - \frac{B_3}{2}\right)\frac{\partial^2 w}{\partial x^2} - B_2\frac{\partial^4 w}{\partial x^4}\\
P_2 = \frac{B_3}{2}\cdot\frac{\partial\mathrm{w}}{\partial\mathrm{x}} + \mathrm{B}_2\frac{\partial^3\mathrm{w}}{\partial\mathrm{x}^3}
\end{cases}
\]
The energy velocity of the wave propagation equals the time averaged active energy flux $\langle P\rangle$ over the time averaged total energy $\langle E_{\mathrm{total}}\rangle$, which can be expressed as
\begin{equation}
\textstyle V_{\mathrm{e}} = \frac{\langle P\rangle}{\langle E_{\mathrm{total}}\rangle}
\label{eq:29}
\end{equation}

\subsection{Frequency response analysis}
\label{sec:3.5}
\begin{figure}[H]
\centering
\includegraphics[width=0.55\linewidth]{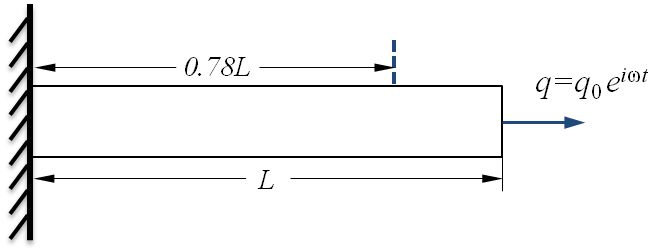}
\caption{A fixed-free micro-rod with loading in free end.}
\label{fig:2}
\end{figure}

As shown in Fig.~\ref{fig:2}, a fix-free rod subject to a harmonic load $q = q_0\mathrm{e}^{\mathrm{i}\omega t}$ at the free end is considered for frequency response function analysis. To be specified here, according to the boundary condition of this case, the frequency normalized parameter is set to be $\omega_k = \sqrt{\frac{E}{\rho}}\cdot\frac{2\kappa}{L}$ to match the first resonant frequency at $\bar{\omega}=1$. The observation point is at $x = 0.78L$. The radius of the cross-section $r = 3a_0$ and the length $L = 8r$.

The solution for the displacement can be given by the superposition of the progressive and retrograde contribution from all wavenumbers ($\bar{k}_1$, $\bar{k}_2$ and $\bar{k}_3$). The general expressions of the dimensionless displacement $\bar{w}$ can be expressed as
\begin{equation}
\bar{w}(\bar{x},\tau) = \left(\tilde{A}\mathrm{e}^{-\mathrm{i}\bar{k}_1\bar{x}} + \tilde{B}\mathrm{e}^{\mathrm{i}\bar{k}_1\bar{x}} + \tilde{C}\mathrm{e}^{-\mathrm{i}\bar{k}_2\bar{x}} + \tilde{D}\mathrm{e}^{\mathrm{i}\bar{k}_2\bar{x}} + \tilde{E}\mathrm{e}^{-\mathrm{i}\bar{k}_3\bar{x}} + \tilde{F}\mathrm{e}^{\mathrm{i}\bar{k}_3\bar{x}}\right)\mathrm{e}^{\mathrm{i}\bar{\omega}\tau}
\label{eq:30}
\end{equation}
From the normalized dispersion relation, which is obtained by the governing equation in Section~\ref{sec:3.2}, all the wavenumbers $\bar{k}_1$, $\bar{k}_2$ and $\bar{k}_3$ on each frequency can be obtained. Then the value of the normalized displacement on one frequency only depends on the amplitudes ($\tilde{A}$, $\tilde{B}$, $\tilde{C}$, $\tilde{D}$, $\tilde{E}$, $\tilde{F}$). Normalizing the harmonic excitation as $\bar{q} = q_0\mathrm{e}^{\mathrm{i}\omega t}/\mu A$, then based on the dimensionless boundary conditions as
\begin{equation}
\begin{cases}
\bar{w}(0,\tau) = 0\quad\text{at}\quad\bar{x} = 0\\
\bar{P}_1(0,\tau) = \frac{2B_1 - B_3}{2\mu AL^2}\cdot\frac{\partial^2\bar{w}}{\partial\bar{x}^2} - \frac{B_2}{\mu AL^4}\cdot\frac{\partial^4\bar{w}}{\partial\bar{x}^4} = 0\qquad\text{at}\qquad\bar{x} = 0\\
\bar{P}_2(0,\tau) = \frac{B_3}{2\mu AL^2}\cdot\frac{\partial\bar{w}}{\partial\bar{x}} + \frac{B_2}{\mu AL^4}\cdot\frac{\partial^3\bar{w}}{\partial\bar{x}^3} = 0\qquad\text{at}\qquad\bar{x} = 0\\
\bar{P}_0(1,\tau) = \frac{E}{\mu}\cdot\frac{\partial\bar{w}}{\partial\bar{x}} + \frac{B_3 - B_1}{\mu AL^2}\cdot\frac{\partial^3\bar{w}}{\partial\bar{x}^3} + \frac{B_2}{\mu AL^4}\cdot\frac{\partial^5\bar{w}}{\partial\bar{x}^5} = \text{-}\ \frac{q_0}{\mu A}\mathrm{e}^{\mathrm{i}\bar{\omega}\tau}\qquad\text{at}\\
\bar{x} = 1\\
\bar{P}_1(1,\tau) = \frac{2B_1 - B_3}{2\mu AL^2}\cdot\frac{\partial^2\bar{w}}{\partial\bar{x}^2} - \frac{B_2}{\mu AL^4}\cdot\frac{\partial^4\bar{w}}{\partial\bar{x}^4} = 0\qquad\text{at}\qquad\bar{x} = 1\\
\bar{P}_2(1,\tau) = \frac{B_3}{2\mu AL^2}\cdot\frac{\partial\bar{w}}{\partial\bar{x}} + \frac{B_2}{\mu AL^4}\cdot\frac{\partial^3\bar{w}}{\partial\bar{x}^3} = 0\qquad\text{at}\qquad\bar{x} = 1
\end{cases}
\label{eq:31}
\end{equation}
These boundary conditions physically means: at the fixed end $\bar{x}=0$, the displacement equals zero, and the higher order forces equal zero; at the loading end $\bar{x}=1$, the produced classical force follows the force equilibrium, and the higher order forces equal zero. The extra kinematic parameters $\frac{\partial\bar{w}}{\partial\bar{x}}$ and $\frac{\partial^2\bar{w}}{\partial\bar{x}^2}$ represent the inner micro-rotation and micro-curvature of the adjacent micro-structures, and the corresponding generalized loads $P_2$ and $P_3$ represent the inner force interactions of the adjacent micro-structures dual to $\frac{\partial\bar{w}}{\partial\bar{x}}$ and $\frac{\partial^2\bar{w}}{\partial\bar{x}^2}$. There are also two kinds of boundary conditions for the enriched model: a loading type and a geometric type. In this case, at $\bar{x}=0$, the possible force produced by the fixed attachment is only classical force, so the higher order force $P_2$ and $P_3$ should be set to zero. At $\bar{x}=1$, $P_0$ follows the force equilibrium on that section, and the higher order force $P_1$ and $P_2$ are set to zero because there is not external higher order force applied on that section.
Following the boundary conditions, six equilibrium with the amplitudes of each progressive and retrograde wave ($\tilde{A}$, $\tilde{B}$, $\tilde{C}$, $\tilde{D}$, $\tilde{E}$, $\tilde{F}$) as variables are established. With some numerical calculation, the amplitude at observation point on each frequency can be obtained.

\section{Wave reflection and the transmission through planar interface based on the enriched model}
\label{sec:4}
\begin{figure}[H]
\centering
\includegraphics[width=0.75\linewidth]{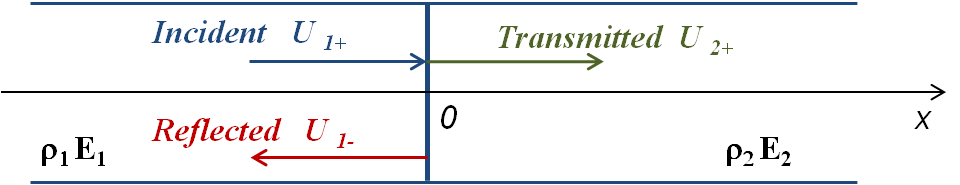}
\caption{Reflection and transmission of longitudinal wave.}
\label{fig:3}
\end{figure}

According to Section~\ref{sec:3.2}, it can be predicted that the body wave modes are different in second strain gradient continua due to the micro-structure effect, and that will lead to significant differences in elastic transmitted and reflected waves. In this section, the reflection and transmission coefficients on a discontinuous surface as well as the energy transmitted ratio are discussed based on the proposed formulation. Fig.~\ref{fig:3} shows rod 1 and rod 2, which are connected together with same circular cross section but different materials.

From the dispersion relation in Sections~\ref{sec:3.2}, we know that 3 modes can be created in second gradient continua. Assuming modes $k_1,k_2,k_3$ can propagate on rod 1 while $k_4,k_5,k_6$ propagate on rod 2, and the incident wave wave $k_1$ vertically inject on the discontinuous surface with amplitude $A_1$, then the displacement caused by incident wave $U_{1+}$ can be expressed as
\begin{equation}
U_{1+} = \tilde{A}_1\cdot\mathrm{e}^{\mathrm{i}(\omega t - k_1x)}
\label{eq:32}
\end{equation}
The incident wave $k_1$ will diffuse out these six wave modes on the discontinuous surface which propagate along the negative $x$ direction and positive $x$ direction respectively. Assuming the amplitudes of the reflected $k_1,k_2,k_3$ wave modes are denoted by $\tilde{B}_1,\tilde{C}_1,\tilde{D}_1$; the amplitudes of transmitted wave modes $k_4,k_5,k_6$ are denoted by $\tilde{B}_2,\tilde{C}_2,\tilde{D}_2$, then the displacement caused by the reflected waves $U_{1-}$ and transmitted waves $U_{2+}$ can be written as (omitting the time component $\mathrm{e}^{\mathrm{i}\omega t}$)
\begin{equation}
\begin{cases}
U_{1-} = \tilde{B}_1\cdot\mathrm{e}^{\mathrm{i}k_1x} + \tilde{C}_1\cdot\mathrm{e}^{\mathrm{i}k_2x} + \tilde{D}_1\cdot\mathrm{e}^{\mathrm{i}k_3x}\\
U_{2+} = \tilde{B}_2\cdot\mathrm{e}^{-\mathrm{i}k_4x} + \tilde{C}_2\cdot\mathrm{e}^{-\mathrm{i}k_5x} + \tilde{D}_2\cdot\mathrm{e}^{-\mathrm{i}k_6x}
\end{cases}
\label{eq:33}
\end{equation}
in which `1' indicates the medium, `$+$', `--' indicates the wave propagation direction. The displacement on the left of interface $U_{\mathrm{L}}$ and the displacement on the right of the interface $U_{\mathrm{R}}$ are given by the superposition of the contribution from all wave modes as
\begin{equation}
\begin{cases}
U_{\mathrm{L}} = \tilde{A}_1\cdot\mathrm{e}^{-\mathrm{i}k_1x} + \tilde{B}_1\cdot\mathrm{e}^{\mathrm{i}k_1x} + \tilde{C}_1\cdot\mathrm{e}^{\mathrm{i}k_2x} + \tilde{D}_1\cdot\mathrm{e}^{\mathrm{i}k_3x}\\
U_{\mathrm{R}} = \tilde{B}_2\cdot\mathrm{e}^{-\mathrm{i}k_4x} + \tilde{C}_2\cdot\mathrm{e}^{-\mathrm{i}k_5x} + \tilde{D}_2\cdot\mathrm{e}^{-\mathrm{i}k_6x}
\end{cases}
\label{eq:34}
\end{equation}
in which `L', `R' indicate the left media and right media. The force and higher order forces on the left of the interface in medium 1 can be expressed with the displacement $U_{\mathrm{L}}$ as
\begin{equation}
\begin{cases}
P_0|_{w=U_{\mathrm{L}}} = EA\frac{\partial U_{\mathrm{L}}}{\partial x} + (B_3 - B_1)\frac{\partial^3 U_{\mathrm{L}}}{\partial x^3} + B_2\frac{\partial^5 U_{\mathrm{L}}}{\partial x^5}\\
P_1|_{w=U_{\mathrm{L}}} = \left(B_1 - \frac{B_3}{2}\right)\frac{\partial^2 U_{\mathrm{L}}}{\partial x^2} - B_2\frac{\partial^4 U_{\mathrm{L}}}{\partial x^4}\\
P_2|_{w=U_{\mathrm{L}}} = \frac{B_3}{2}\cdot\frac{\partial\mathrm{U_L}}{\partial\mathrm{x}} + \mathrm{B}_2\frac{\partial^3\mathrm{U_L}}{\partial\mathrm{x}^3}
\end{cases}
\label{eq:35}
\end{equation}
While the force and higher order forces on the right of the interface in medium 2 can also be expressed with the displacement $U_{\mathrm{R}}$ as
\begin{equation}
\begin{cases}
P_0|_{w=U_{\mathrm{R}}} = EA\frac{\partial U_{\mathrm{R}}}{\partial x} + (B_3 - B_1)\frac{\partial^3 U_{\mathrm{R}}}{\partial x^3} + B_2\frac{\partial^5 U_{\mathrm{R}}}{\partial x^5}\\
P_1|_{w=U_{\mathrm{R}}} = \left(B_1 - \frac{B_3}{2}\right)\frac{\partial^2 U_{\mathrm{R}}}{\partial x^2} - B_2\frac{\partial^4 U_{\mathrm{R}}}{\partial x^4}
P_2|_{w=U_{\mathrm{R}}} = \frac{B_3}{2}\cdot\frac{\partial\mathrm{U_R}}{\partial\mathrm{x}} + \mathrm{B}_2\frac{\partial^3\mathrm{U_R}}{\partial\mathrm{x}^3}
\end{cases}
\label{eq:36}
\end{equation}
Assuming the interface does not dissipate energy, and the mass of the rod between $x=0^+$ and $x=0^-$ is infinitesimal, the force equilibrium can be developed at $x=0$. Amplitudes of each transmitted waves and reflected waves can be estimated based on the continuity of displacement, the first derivative of the displacement, and the second derivative of the displacement along with the equilibrium of the force and higher order forces as
\begin{equation}
\textstyle \left.\left[w,\frac{\partial w}{\partial x},\frac{\partial^2 w}{\partial x^2},P_0,P_1,P_2\right]\right|_{w=U_L} = \left.\left[w,\frac{\partial w}{\partial x},\frac{\partial^2 w}{\partial x^2},P_0,P_1,P_2\right]\right|_{w=U_R}
\label{eq:37}
\end{equation}
In the following numerical calculation, the amplitudes of the reflected waves and transmitted waves can be calculated and expressed with $\tilde{A}_1$, thus the reflection coefficients and transmission coefficients for wave $k_1$, $k_2$ and $k_3$ are presented in the form of amplitude ratios as follows.
\begin{equation}
\textstyle R_1 = \frac{\tilde{B}_1}{\tilde{A}_1},\ R_2 = \frac{\tilde{C}_1}{\tilde{A}_1},\ R_3 = \frac{\tilde{D}_1}{\tilde{A}_1},T_1 = \frac{\tilde{B}_2}{\tilde{A}_1},\ T_2 = \frac{\tilde{C}_2}{\tilde{A}_1},\ T_3 = \frac{\tilde{D}_2}{\tilde{A}_1}.
\label{eq:38}
\end{equation}
$R_1$, $R_2$ and $R_3$ are the reflection coefficients for wave $k_1$, $k_2$ and $k_3$; $T_1$, $T_2$ and $T_3$ denote the transmission coefficients for wave $k_4$, $k_5$ and $k_6$.
\begin{table}[H]
\small
\centering
\caption{High-order material constants for aluminum.\cite{r40}}
\label{tab:1}
\begin{tabular}{ccccccc}
\toprule
$\begin{array}{l}{a_1}( {\rm{eV/A }} )\\0.1407\end{array}$&
$\begin{array}{l}{a_2}( {\rm{eV/A }} )\\0.0027\end{array}$&
$\begin{array}{l}{a_3}( {\rm{eV/A }} )\\-0.0083\end{array}$&
$\begin{array}{l}{a_4}( {\rm{eV/A }} )\\0.0966\end{array}$&
$\begin{array}{l}{a_5}( {\rm{eV/A }} )\\0.2584\end{array}$& & \\
$\begin{array}{l}{b_1}( {\rm{eV/A }} )\\0.7927\end{array}$&
$\begin{array}{l}{b_2}( {\rm{eV/A }} )\\0.0644\end{array}$&
$\begin{array}{l}{b_3}( {\rm{eV/A }} )\\-0.1943\end{array}$&
$\begin{array}{l}{b_4}( {\rm{eV/A }} )\\-0.0009\end{array}$&
$\begin{array}{l}{b_5}( {\rm{eV/A }} )\\-0.0009\end{array}$&
$\begin{array}{l}{b_6}( {\rm{eV/A }} )\\16.1566\end{array}$&
$\begin{array}{l}{b_7}( {\rm{eV/A }} )\\48.5291\end{array}$\\
$\begin{array}{l}{c_1}( {\rm{eV/A }} )\\0.5041\end{array}$&
$\begin{array}{l}{c_2}( {\rm{eV/A }} )\\0.3569\end{array}$&
$\begin{array}{l}{c_3}( {\rm{eV/A }} )\\0.1782\end{array}$&
& & & \\
\bottomrule
\end{tabular}
\end{table}
According to Section~\ref{sec:3.4}, energy flux in second strain gradient continua is coupled with the classical force and the higher order forces, hence the reflected energy flux and transmitted energy flux based on the proposed SSG model can be expressed out as
\begin{equation}
\textstyle J_{\mathrm{reflection}} = -\frac{1}{2}Re\left.\left[P_0\cdot\dot{U}_{1-}^* + P_1\cdot\left(\frac{\partial\dot{U}_{1-}}{\partial x}\right)^* + P_2\cdot\left(\frac{\partial^2\dot{U}_{1-}}{\partial x^2}\right)^*\right]\right|_{w=U_{1-}}
\label{eq:39}
\end{equation}
and
\begin{equation}
\textstyle J_{\mathrm{transmission}} = -\frac{1}{2}Re\left.\left[P_0\cdot\dot{U}_{2+}^* + P_1\cdot\left(\frac{\partial\dot{U}_{2+}}{\partial x}\right)^* + P_2\cdot\left(\frac{\partial^2\dot{U}_{2+}}{\partial x^2}\right)^*\right]\right|_{w=U_{2+}}
\label{eq:40}
\end{equation}
$P_0$, $P_1$ and $P_2$ in the expressions are in dependence on the displacement in each media. The energy distributed in transmitted and reflected waves can be achieved by the calculating the ratio between the transmitted energy as well as the reflected energy with the incident energy flux.

Until here, the formulation of the governing equation and the associated boundary conditions of the SSG theory based rod model are fully established. The characterization of wave dispersion relation and modal density, the frequency response of bounded SSG theory based rod, the energy flux in second gradient continua and wave reflection and transmission coefficients on the discontinuous surface are all completed.

\section{Numerical application and discussions}
\label{sec:5}
In this section, cases are studied to illustrate enriched wave properties based on SSG theory model. The structure of the rod is shown as in Fig.~\ref{fig:1}\cite{r40}. The material is assumed to be aluminum with $\mu=26\,\mathrm{GPa}$, and the higher order material constant values are given in Table~\ref{tab:1} with lattice parameter $a_0=4.04\mathrm{A}$. The rod has a circular cross section with radius $r=3a_0$, and length $L=8r$.

\subsection{Dispersion curves}
\label{sec:5.1}
\begin{figure}[H]
\centering
\includegraphics[width=0.95\linewidth]{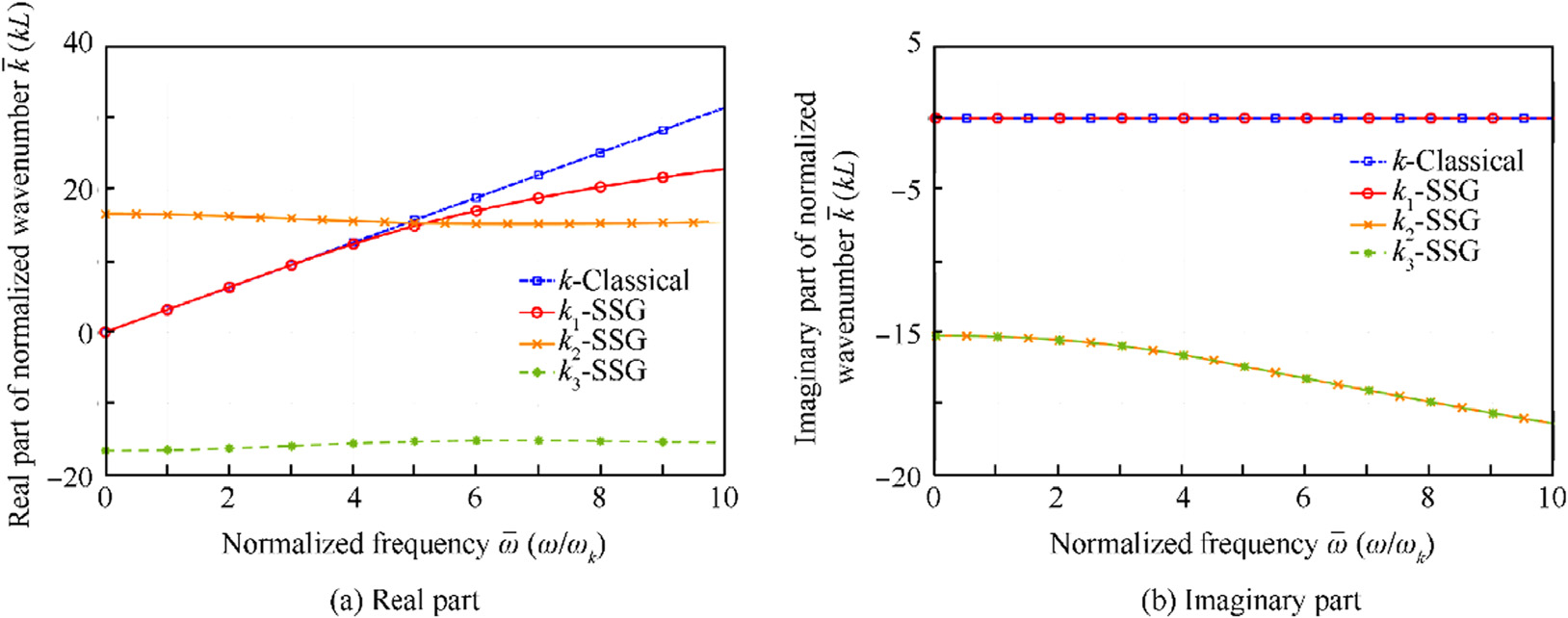}
\caption{Dispersion relation of positive-going waves $k^{+}(\omega)$.}
\label{fig:4}
\end{figure}

\begin{figure}[H]
\centering
\includegraphics[width=0.95\linewidth]{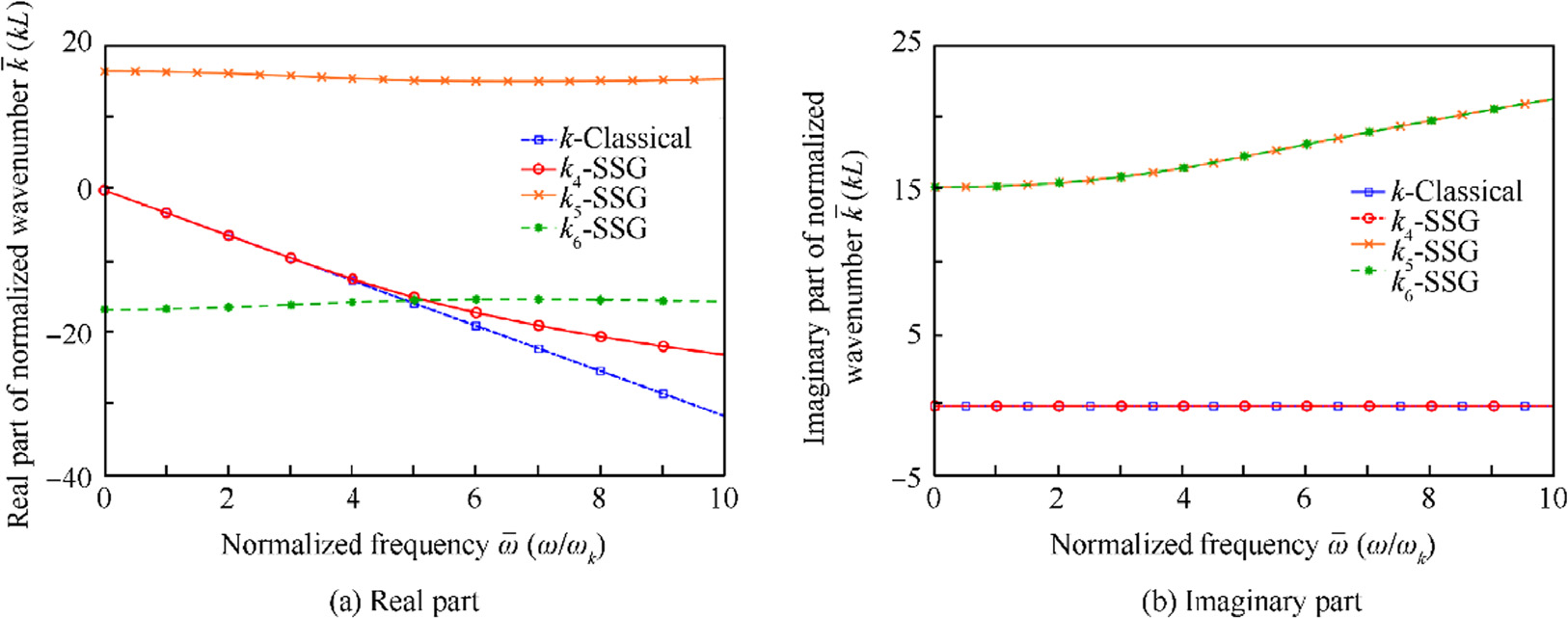}
\caption{Dispersion relation of negative-going waves $k^{-}(\omega)$.}
\label{fig:5}
\end{figure}

Employing the parameters above for the dispersion Eq.~(18), one arrives six solutions, which indicates 3 positive going modes and 3 negative going modes on each frequency. The results are shown in Fig.~\ref{fig:4} and Fig.~\ref{fig:5}.
Fig.~\ref{fig:4}. and Fig.~\ref{fig:5}. illustrate the dispersion relation of positive going waves and negative going waves, respectively. It should be noticed here, the wavenumber $k_3$, with negative imaginary part and negative real part is considered as a positive-going wave, and $k_6$ with positive imaginary part and positive real part as a negative-going wave. In a periodic waveguide with unit-cell's dimension $d$, $k$ is $2m\pi/d$ -periodic, the positive- or negative-going waves can have arbitrary Real$(k)$ sign. The direction of propagation is defined by the Imag$(k)$, which corresponds to the wave attenuation.

Fig.~\ref{fig:4} has two subfigures, Fig.~\ref{fig:4} (a) and Fig.~\ref{fig:4} (b), which display the real part and imaginary part of the dimensionless wavenumbers for the positive-going waves in frequency range $[0,10\,\omega_k]$. Different from the classical theory results, 3 modes $(k_1,k_2,k_3)$ are predicted by the enriched model, in which the non-classical longitudinal wave $k_1$ is propagating in a dispersive manner. Real part of $k_1$ is smaller than classical one and the gap widens with the increasing frequency. The other two wave $k_2$ and $k_3$ appear exclusively in SSG theory model, and they are both evanescent waves. These results agree with the Refs. \citetext{\citenum{r32},\citenum{r33},\citenum{r44}}.

\begin{figure}[H]
\centering
\includegraphics[width=0.7\linewidth]{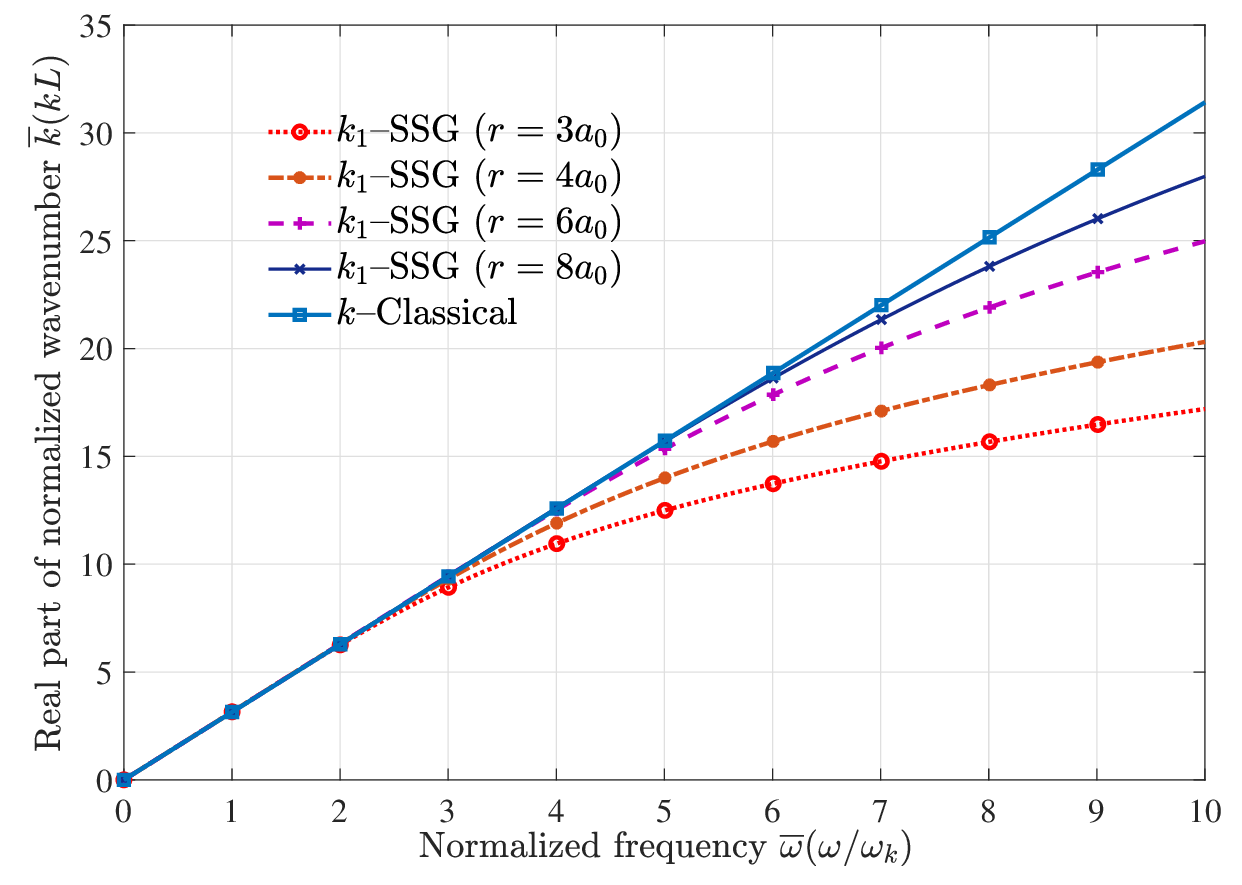}
\caption{Wavenumber changing with rod dimension.}
\label{fig:6}
\end{figure}

In order to study the scale-effect of the dispersive feature, wavenumbers of mode $k_1$ propagating in rod structures of different size are investigated. The results are illustrated in Fig.~\ref{fig:6}.

\begin{figure}[H]
\centering
\includegraphics[width=0.7\linewidth]{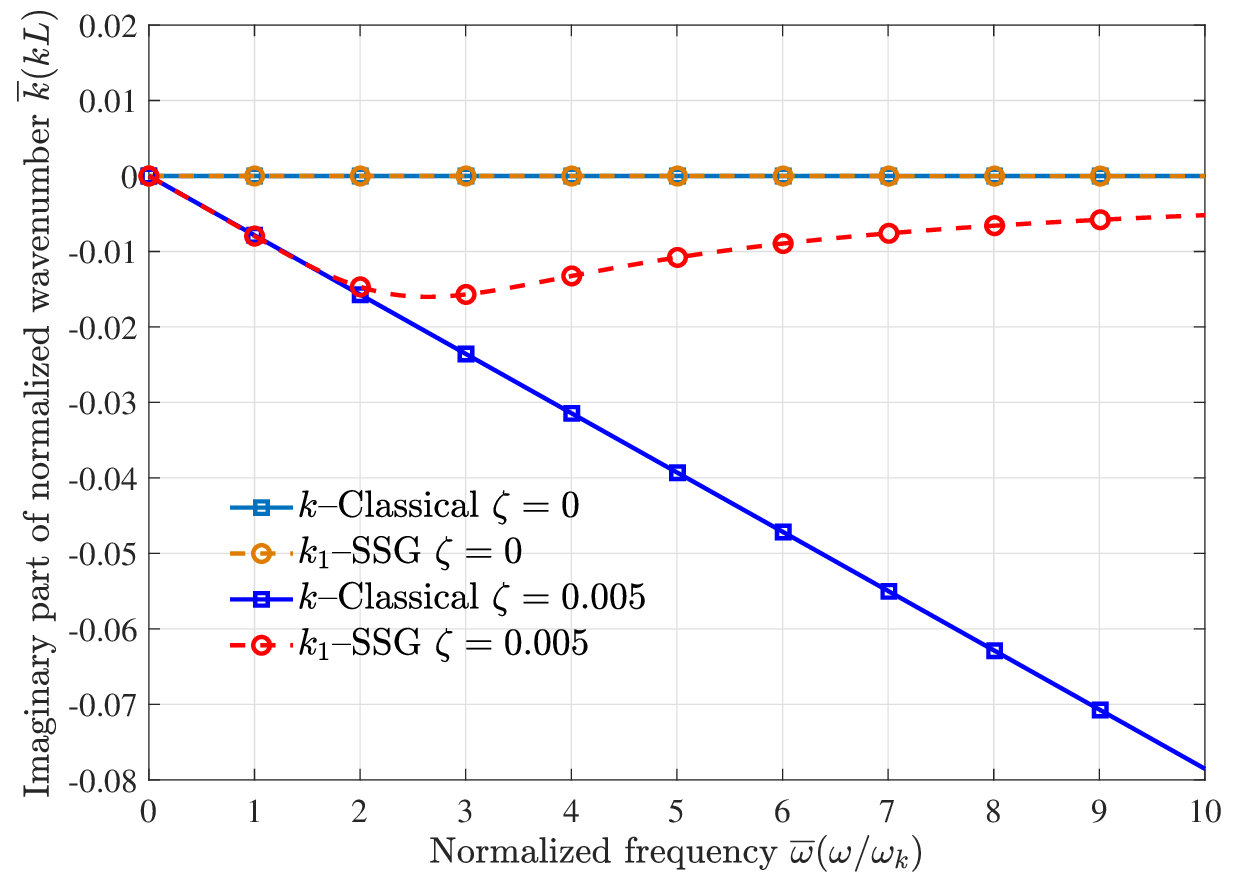}
\caption{Wavenumber changing with damping.}
\label{fig:7}
\end{figure}

As dimension goes up, length to radius ratio of the structure is fixed as $L=5r$, hence the resulting dispersion curve of classical model is linear correlated with angular frequency. Whereas the dispersion curve of SSG theory is observed to keep changing and approaching the classical result as the radius $r$ goes from $3a_0$ to $8a_0$. When the radius increase, the structural characteristic length grows longer. Therefore, this phenomenon reflects that the influence of micro-structure interaction weakens as the structural characteristic length goes up.
The above outcomes are all calculated without energy dissipation. However, in practical applications, there are always damping in the system. Energy dissipation can significantly influence wave propagation features. Assuming the damping is viscous, then $E=52\,(1+\mathrm{i}\zeta)\,\mathrm{GPa}$ can be utilized with $\zeta$ indicating the loss factor of the system. To analyze the influence of damping for the proposed enriched model, the dispersion curves with $\zeta=0.005$ is shown in Fig.~\ref{fig:7}.

In relatively low frequency range $[0,2\omega_k]$, the results of SSG model and classical model are similar. In higher frequency range $[2\omega_k,10\omega_k]$, wave number is evidently influenced for the non-classical wave $k_1$, as the absolute value of imaginary parts decrease and draw near to zero. Energy dissipation through viscous damping for non-classical longitudinal wave $k_1$ become negligible compared with the classical model.
\subsection{Modal density}
\label{sec:5.2}
\begin{figure}[H]
\centering
\includegraphics[width=0.7\linewidth]{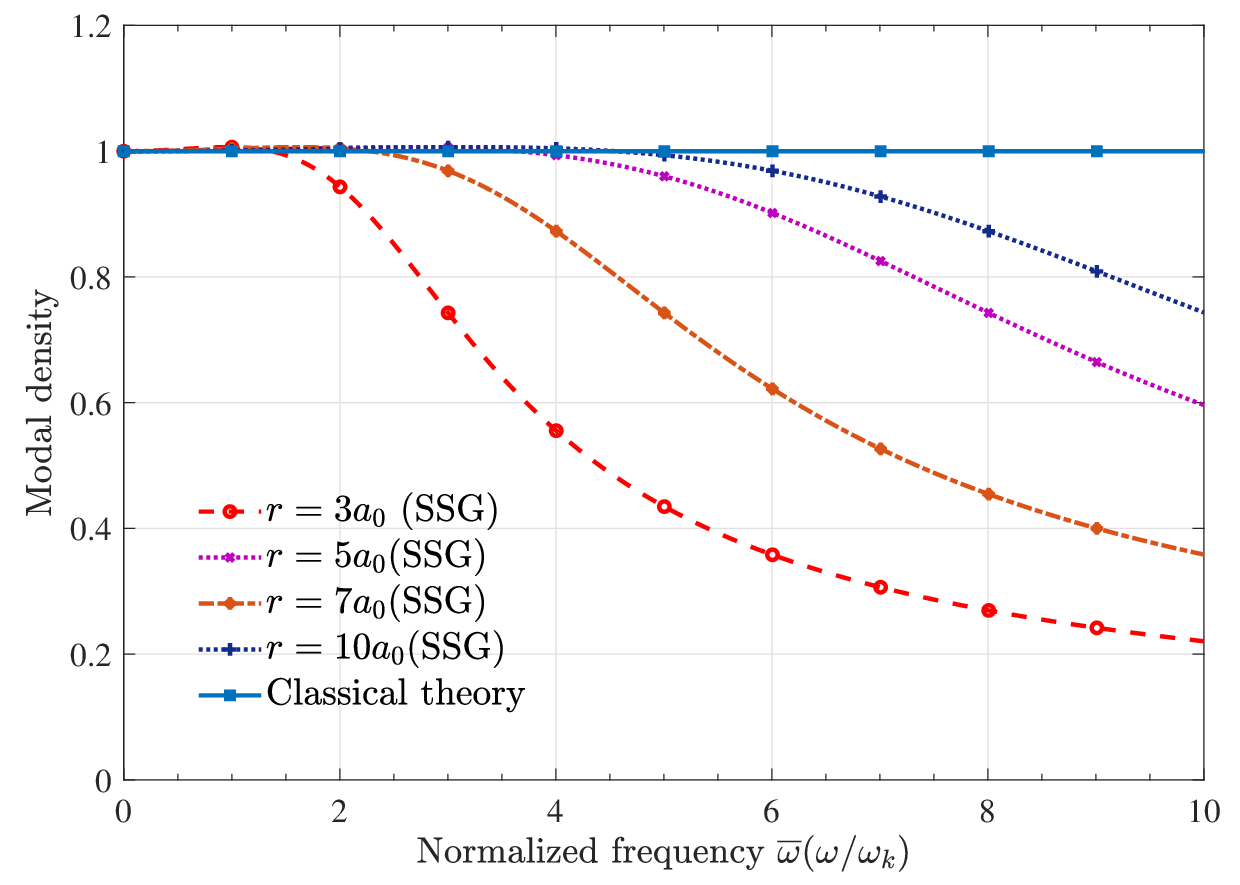}
\caption{Modal density.}
\label{fig:8}
\end{figure}

Fig.~\ref{fig:8} depicts modal density of non-classical wave $k_1$ compared with classical result, which keeps still in frequency domain. The influence of micro-structure effect is quite obvious as the non-classical result is size and frequency dependent. Modal density of $k_1$ decrease significantly compared with classical results. The contrast becomes more serious as the model's radius decreasing gradually to $r=3a_0$. The present SSG theory model can clearly capture the micro-structure effect. This phenomenon is similar to the observation in the case of wavenumber characterization.
\subsection{Energy velocity}
\label{sec:5.3}
In Section~\ref{sec:3.4}, we established the energy flux expression for SSG theory based rod model. The validation of energy flux expression is important for the following researches which are developed from energy equilibrium. In this section, energy velocity $V_{\mathrm{e}}$ and group velocity $c_{\mathrm{g}}$ are illustrated for non-classical wave $k_1$. If the formulation Eq.~(\ref{eq:27}) and Eq.~(\ref{eq:28}) are validated, group velocity $c_{\mathrm{g}}$ should be identical to the energy velocity $V_{\mathrm{e}}$ without damping in the system.

\begin{figure}[H]
\centering
\includegraphics[width=0.7\linewidth]{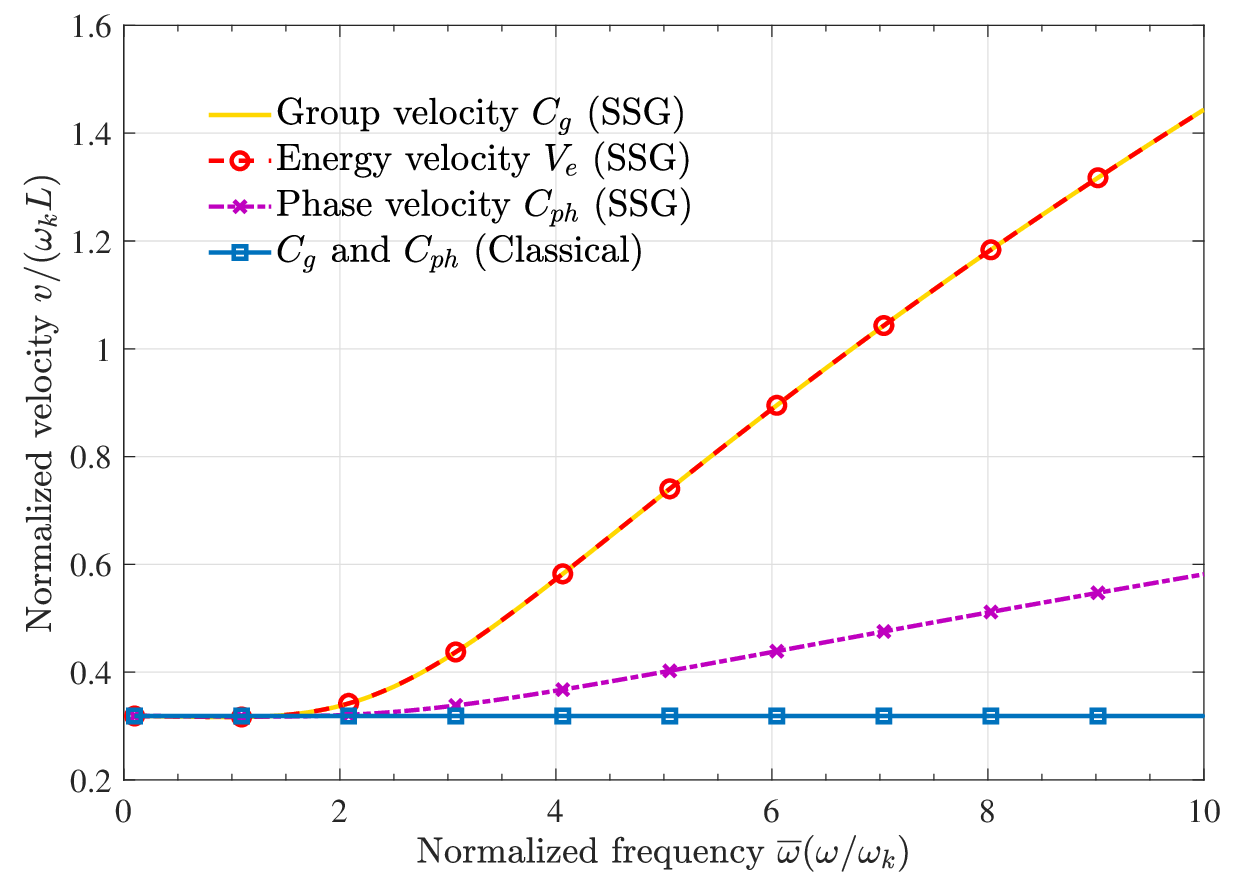}
\caption{Energy velocity of longitudinal wave.}
\label{fig:9}
\end{figure}

From Fig.~\ref{fig:9}, as expected, the formulation for energy flux in SSG theory based rod model are proved to be correct with $c_{\mathrm{g}}=V_{\mathrm{e}}$ in the whole frequency range. In addition, group velocity $c_{\mathrm{g}}$ and phase velocity $c_{\mathrm{ph}}$ of $k_1$ in SSG theory model are clearly different from each other, which confirms the dispersive property of non-classical $k_1$.
\subsection{Frequency response function}
\label{sec:5.4}
For frequency response analysis, one harmonic force $q=0.05\mu A\mathrm{e}^{\mathrm{i}\omega t}$ is applied on free end of the structure. Based on the previous formulation, the amplitude of each mode at observation point $\bar{x}=0.78$ are calculated out on each frequency, and the displacement magnitudes are obtained by the superposition of each mode.

\begin{figure}[H]
\centering
\includegraphics[width=0.7\linewidth]{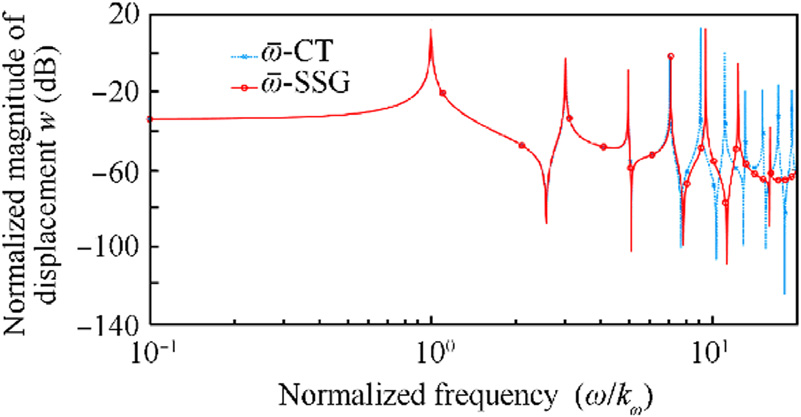}
\caption{Frequency response function at $x=0.78L$.}
\label{fig:10}
\end{figure}

Fig.~\ref{fig:10} displays the normalized displacement magnitudes at observation point based on SSG theory and classical theory. We notice that resonances can be well predicted by both models. In lower frequency range, the results of these two models match well, but in higher frequency, the values of SSG theory shift to higher frequencies compared to classical ones, and less resonance peaks are observed in frequency range $[0.1\omega_k,10\omega_k]$ (7 for SSG theory and 10 for classical theory). Due to the influence of micro-structure interactions, wave propagation is significantly affected, vibration energy is not only conserved in the propagating longitudinal wave, but also converted into the other two evanescent modes. This can be the explanation for the reduction of resonance peaks. In higher frequency, micro-structure effect could cause more energy conversion into the extra evanescent waves.
\begin{table}[H]
\small
\centering
\caption{High-order material constants for copper.\cite{r40}}
\label{tab:2}
\begin{tabular}{ccccccc}
\toprule
$\begin{array}{l}{a_1}( {\rm{eV/A }} )\\0.1833\end{array}$&
$\begin{array}{l}{a_2}( {\rm{eV/A }} )\\0.0103\end{array}$&
$\begin{array}{l}{a_3}( {\rm{eV/A }} )\\0.0010\end{array}$&
$\begin{array}{l}{a_4}( {\rm{eV/A }} )\\0.0717\end{array}$&
$\begin{array}{l}{a_5}( {\rm{eV/A }} )\\0.1891\end{array}$& & \\
$\begin{array}{l}{b_1}( {\rm{eV/A }} )\\0.6612\end{array}$&
$\begin{array}{l}{b_2}( {\rm{eV/A }} )\\0.0663\end{array}$&
$\begin{array}{l}{b_3}( {\rm{eV/A }} )\\-0.2062\end{array}$&
$\begin{array}{l}{b_4}( {\rm{eV/A }} )\\-0.0015\end{array}$&
$\begin{array}{l}{b_5}( {\rm{eV/A }} )\\-0.0015\end{array}$&
$\begin{array}{l}{b_6}( {\rm{eV/A }} )\\12.6254\end{array}$&
$\begin{array}{l}{b_7}( {\rm{eV/A }} )\\37.9402\end{array}$\\
$\begin{array}{l}{c_1}( {\rm{eV/A }} )\\0.8448\end{array}$&
$\begin{array}{l}{c_2}( {\rm{eV/A }} )\\0.5732\end{array}$&
$\begin{array}{l}{c_3}( {\rm{eV/A }} )\\0.3465\end{array}$&
& & & \\
\bottomrule
\end{tabular}
\end{table}
\subsection{Reflection and transmission coefficients}
\label{sec:5.5}
Numerical calculation for the reflected and transmitted wave at the discontinuous surface through different SSG theory based structures is conducted in this section. As shown in Fig.~\ref{fig:3}, rod 1 and rod 2 are connected with same cross section radius $r=3a_0$. Rod 1 is made of aluminum and the material constants have been illustrated in previous section. Rod 2 is copper with $\mu=40\mathrm{GPa}$, and the higher order material constants are shown in Table~\ref{tab:2}\cite{r40}.

Substitute these values into the formulations in Section~\ref{sec:4}, we can easily achieve the reflection and transmission coefficients as well as the energy transmission situation.

\begin{figure}[H]
\centering
\subfigure[Transmission coefficients]{\includegraphics[width=0.42\linewidth]{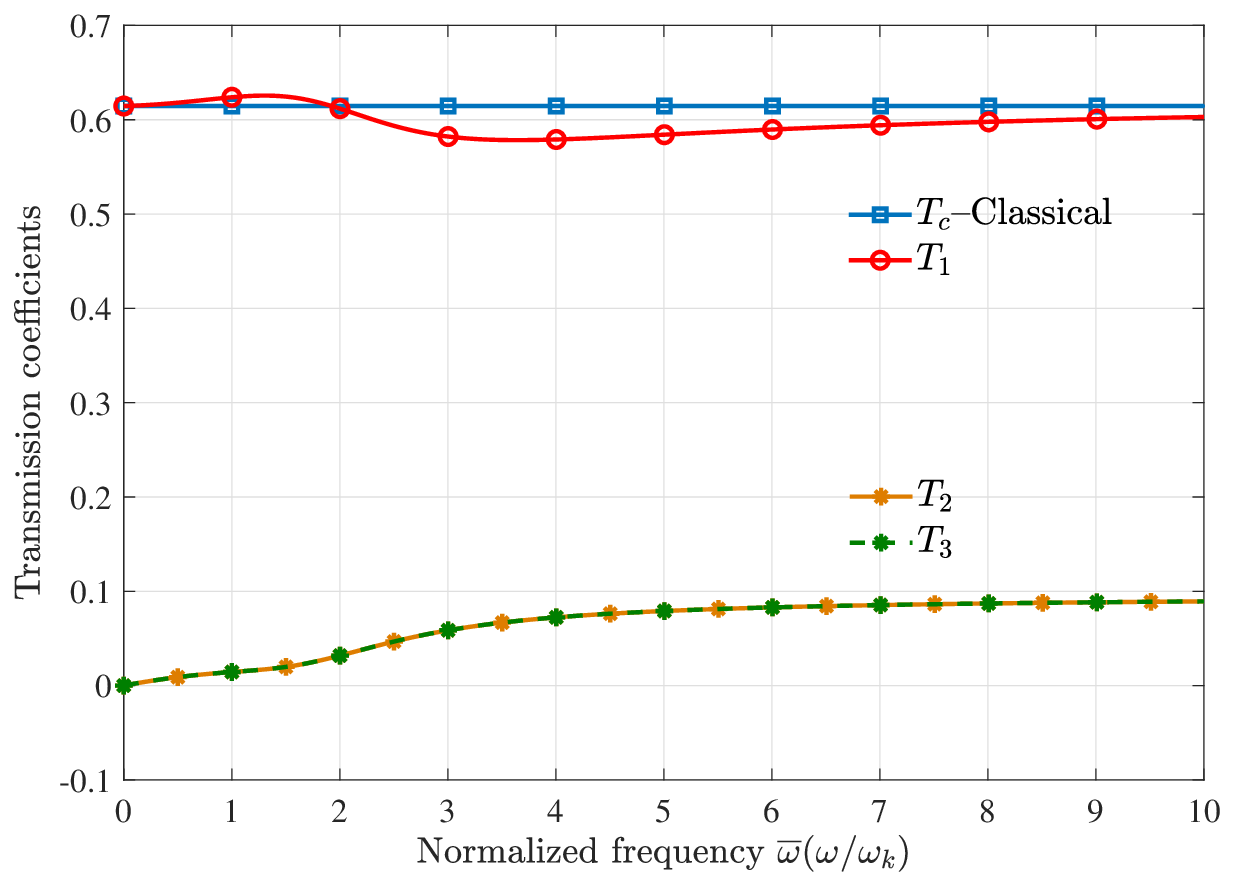}}\hspace{1em}
\subfigure[Reflection coefficients]{\includegraphics[width=0.42\linewidth]{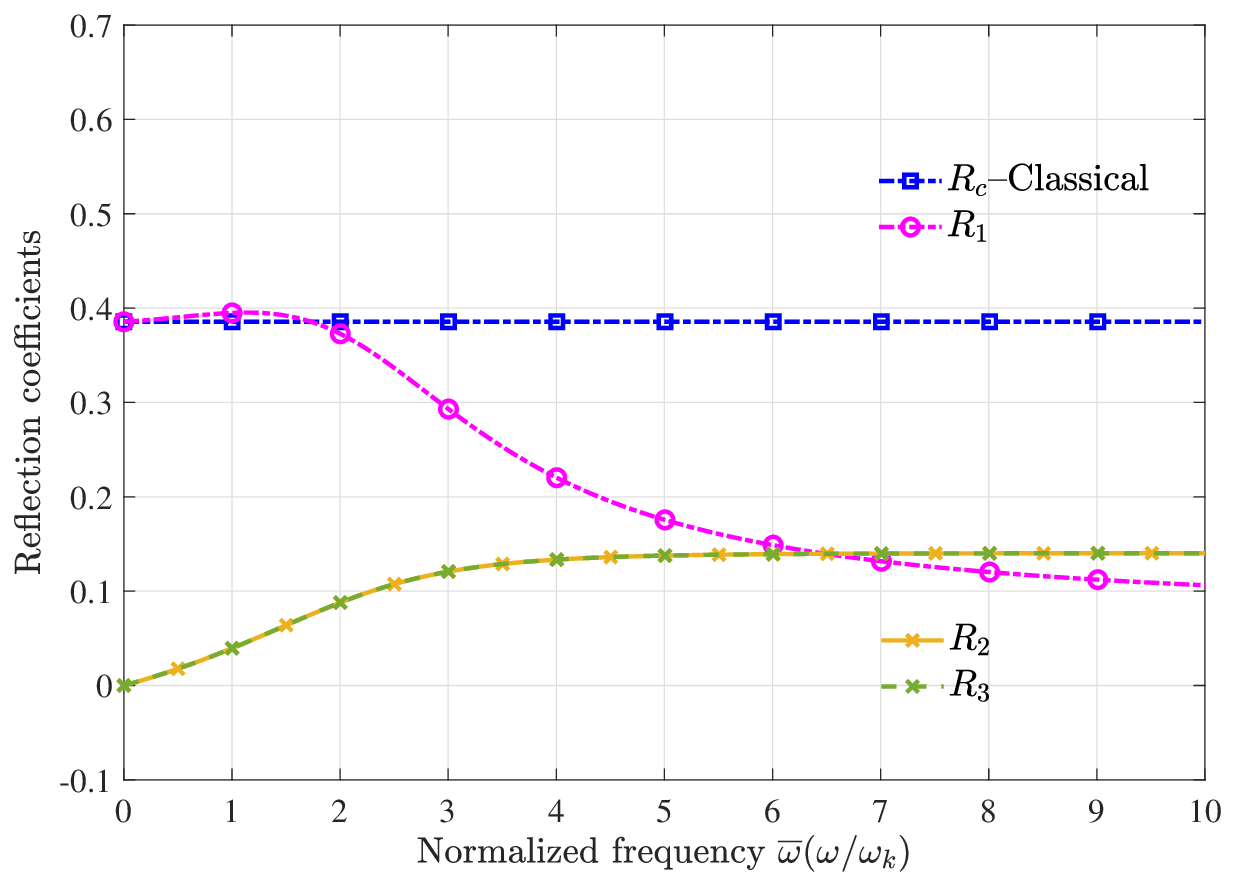}}
\caption{Transmission and reflection coefficients.}
\label{fig:11}
\end{figure}

Fig.~\ref{fig:11} depicts the transmission and reflection coefficients based on SSG theory and classical theory. Compared with the classical transmission coefficient $T_{\mathrm{c}}$ and reflection coefficients $R_{\mathrm{c}}$ which stay still in whole frequency range, the SSG theory results are frequency dependent and change significantly especially for $R_1$. In higher frequency, $R_1$ decrease dramatically compared with the classical result, while $T_1$ changes slightly. Transmission and reflection coefficients $T_2$, $T_3$, $R_2$ and $R_3$ all grow up as frequency goes up.

\begin{figure}[H]
\centering
\begin{minipage}[t]{0.45\linewidth}
\centering
\includegraphics[width=\linewidth]{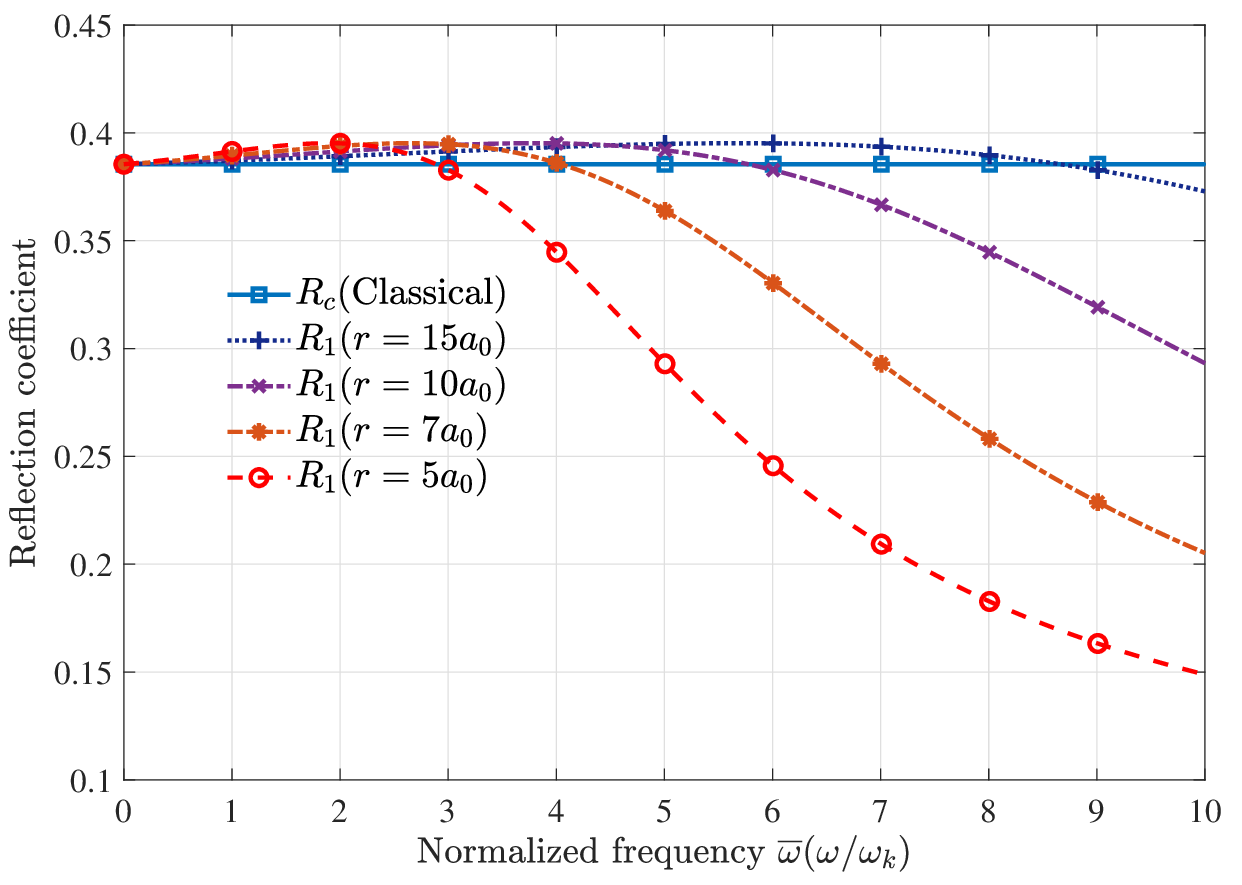}
\caption{Size effect of reflection coefficients.}
\label{fig:12}
\end{minipage}%
\hfill
\begin{minipage}[t]{0.45\linewidth}
\centering
\includegraphics[width=\linewidth]{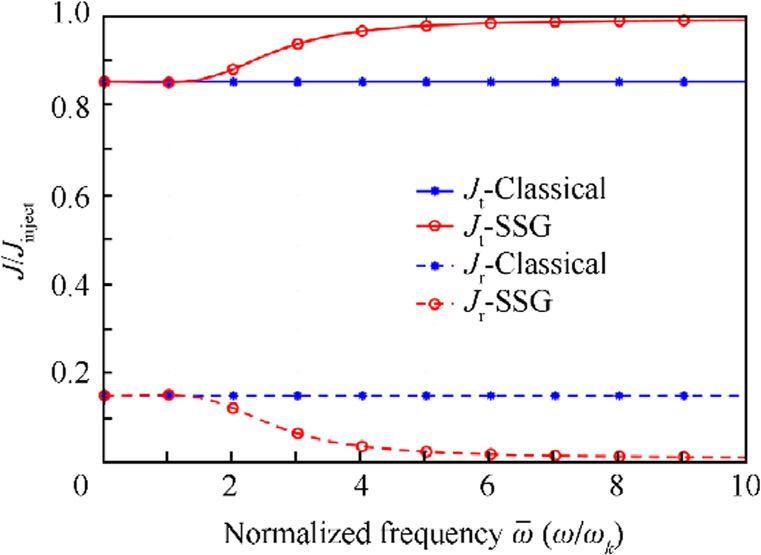}
\caption{Energy transmission and reflection.}
\label{fig:13}
\end{minipage}
\end{figure}

In order to verify the size effect for the wave reflection features, reflection coefficients of wave $k_1$ propagating in these two SSG theory based rods with increasing dimensions are analyzed and the results are illustrated in Fig.~\ref{fig:12}. We can see the reflection coefficient of $k_1$ increase and approach to the classical result as the cross section radius $r$ increasing from $5a_0$ to $15a_0$.

Based on the previous formulation, it is admitted that the energy converted to the extra evanescent waves $k_2$ and $k_3$ leads to great differences in energy transmission. To achieve more information, we intended to calculate the energy conserved in all three modes separately, but it is impossible due to the strong coupling between these three modes refer to energy flux. Therefore, in order to investigate the energy transmission ratio in SSG theory model, Fig.~\ref{fig:13} is plotted.
Fig.~\ref{fig:13} shows the distribution of transmitted energy and reflected energy conserved in all 3 modes. We notice that even though the energy transmitted ratio with SSG theory formulation is increasing and reflected ratio is decreasing, the summation of them is still `1', which indicates the previous formulations are valid. With the enriched model, energy is predicted to transmit more and reflect less. Combined with Fig.~\ref{fig:11} (a), we noted that that the transmitted energy diffuse quite large percentage to wave modes $k_2$ and $k_3$, so $T_1$, the transmission coefficient of wave $k_1$ is actually decreasing. Conversely, the total reflected energy decreases slightly, but with considering the energy diffusion into $k_2$ and $k_3$, the reflection coefficient for wave $k_1$ decreases dramatically.

\section{Conclusions}
\label{sec:6}
Wave propagation in SSG theory based media with consideration of the heterogeneity caused by micro-structure effect is different from the conventional one. In order to analyze this special behavior, we established an enriched model based on Mindlin's Second Strain Gradient (SSG) theory. This formulation allows the micro-structure's effect to be captured by considering the higher order strains and the heterogeneity to be described in the frame of continuum mechanics. From the numerical calculations, one can see that the proposed SSG theory based rod model is effective in predicting the non-classical dispersive behaviors in complex media, and that results are in good agreement with some literature.\citetext{\citenum{r32},\citenum{r33},\citenum{r44}} Apart from the dispersion characteristics, this research also highlighted a number of interesting features of the modal density, the energy flux and the resulting forced response of the rod. Wave transmission and reflection between two different waveguides in second gradient material was also investigated. Some conclusions can be drawn from the numerical calculation and discussions as follows:
\begin{enumerate}
\item[(1)] There are three waves in second gradient rod, one is the propagating non-classical longitudinal wave $k_1$ which exhibits dispersive feature, and the dispersive behavior become more prominent in higher frequency. The other two waves $k_2$ and $k_3$ are both evanescent. In higher frequency, energy dissipation through viscous damping for wave $k_1$ become negligible compared with classical result. As dimension goes up, micro-structural effect is weaken for the reason that the length of inner heterogeneity is less comparable to the increasing structural characteristic length, and wave $k_1$ become less dispersive.
\item[(2)] The proposed formulation of the energy flux is validated, as the energy $V_{\mathrm{e}}$ and group $c_{\mathrm{g}}$ velocities of wave $k_1$ are proved to be identical in all the frequency range. Energy flux is coupled with one classical force and two higher order forces, which are respectively the resultants of classical and higher order stresses through the surface of the section.
\item[(3)] The statistical value of the modal density for the non-classical longitudinal wave $k_1$ decreases drastically compared with classical results. This is due to the influence of micro-structure. This phenomenon becomes more pronounced as the model characteristic size decreases. From the FRF analysis, the resonant frequencies are different, the values of SSG theory shift to higher frequencies compared to classical ones. These unusual behaviors can be explained with different energy transfer through the extra evanescent waves $k_2$ and $k_3$ especially in higher frequency range when the wavelength and the structural characteristic length are in the same order.
\item[(4)] Wave reflection and transmission on a discontinuous surface based on the enriched model are highly different from the classical theory ones. Vibration energy of the incident wave is predicted to transmit more and reflect less, while the transmitted and reflected energy is conserved in all three modes. Due to the influence of modes $k_2$ and $k_3$, energy attenuation increases to some extent. Therefore, the transmission coefficient of the propagating wave $k_1$ decrease slightly and the reflection coefficient decreases drastically. The higher the frequency becomes, the greater the impact of micro-structure.
\end{enumerate}

\section*{Acknowledgements}
This work was supported by the LabEx CeLyA (Centre Lyonnais d'Acoustique, ANR-10-LABX-0060) of Universit\'e de Lyon.

\end{document}